\documentclass[12pt]{extarticle}
\usepackage[utf8]{inputenc}
\usepackage[margin=1in]{geometry}
\usepackage{mathtools}
\usepackage{booktabs}
\usepackage{amsmath}
\usepackage{appendix}
\usepackage{graphicx}
\usepackage[para]{threeparttable}
\usepackage[round]{natbib}
\usepackage{setspace}
\usepackage{lineno}
\usepackage{lipsum}
\usepackage{url}
\usepackage{hyperref}
\usepackage{adjustbox}
\usepackage{float}
\usepackage{tikz}
\usepackage{amsmath}
\usepackage{mathabx}
\usepackage{hyperref}
\usepackage{multirow}
\usepackage{listings}
\usepackage{pdflscape}
\usepackage{caption}
\usepackage{subcaption}

\makeatletter
\renewcommand*\env@matrix[1][\arraystretch]{%
  \edef\arraystretch{#1}%
  \hskip -\arraycolsep
  \let\@ifnextchar\new@ifnextchar
  \array{*\c@MaxMatrixCols c}}
\makeatother

\usetikzlibrary{decorations.pathreplacing}
\usetikzlibrary{arrows}

\begin{document}

\title{\Large{\textbf{Practical considerations for sandwich variance estimation in two-stage regression settings}}}
\author{Lillian A. Boe$^{1,*}$, Thomas Lumley$^{2}$, and 
Pamela A. Shaw$^{3}$ \\
$^{1}$Department of Biostatistics, Epidemiology, and Informatics, \\
University of Pennsylvania Perelman School of Medicine, \\
Philadelphia, Pennsylvania, United States
\\
$^{2}$Department of Statistics, Faculty of Science, University of Auckland, \\
Auckland, New Zealand \\
$^{3}$Biostatistics Unit, Kaiser Permanente Washington Health Research Institute \\
Seattle, Washington, United States \\
\textit{*email}: boel@pennmedicine.upenn.edu}

\date{}
\maketitle


\begin{abstract}
{We present a practical approach for computing the sandwich variance estimator in two-stage regression model settings. As a motivating example for two-stage regression, we consider regression calibration, a popular approach for addressing covariate measurement error. The sandwich variance approach has been rarely applied in regression calibration, despite that it requires less computation time than popular resampling approaches for variance estimation, specifically the bootstrap. This is likely due to requiring specialized statistical coding.  In practice, a simple bootstrap approach with Wald confidence intervals is often applied, but this approach can yield confidence intervals that do not achieve the nominal coverage level.  We first outline the steps needed to compute the sandwich variance estimator. 
We then develop a convenient method of computation in R for sandwich variance estimation, which leverages standard regression model outputs and existing R functions and can be applied in the case of a simple random sample or complex survey design. We use a simulation study to compare the performance of the sandwich to a resampling variance approach for both data settings. Finally, we further compare these two variance estimation approaches for data examples from the Women’s Health Initiative (WHI) and Hispanic Community Health Study/Study of Latinos (HCHS/SOL).}
\end{abstract}
Keywords: Bootstrap, measurement error, regression calibration, sandwich variance, robust variance, two-stage \\~\

\section{Introduction} 
Two stage regression models arise in several settings in epidemiology and statistics, including those requiring correction for covariate measurement error, mediation analysis, and using instrumental variables in causal inference \citep{keogh20,baron1986moderator,baiocchi2014instrumental}. Typically, plug-in estimates for nuisance parameters are obtained in stage 1 and an outcome model reliant on these plug-in estimates is fit in stage 2. As a motivating example, we consider regression calibration, a popular analysis approach for correcting biases in regression coefficients induced by covariate measurement error \citep{shaw2018citation,prentice1982covariate}. When regression calibration is applied, additional steps are required for variance estimation for the outcome model coefficients to account for the uncertainty in the estimated, error-adjusted exposure. Usual standard errors obtained from the outcome model are generally too small, resulting in overly narrow confidence intervals. 

There are a few approaches for variance estimation in two-stage regression. We consider settings where the data are either from a simple random sample (SRS) from the population or a complex survey sampling design. For the SRS case, two common approaches are a bootstrap variance estimator \citep{efron1979bootstrap} or a sandwich variance estimator obtained by stacking the calibration and outcome model estimating equations \citet{boos2013essential}. In practice, the bootstrap estimator is often implemented because it is fairly simple to apply in standard software \citep{keogh20}, but Wald confidence intervals constructed using bootstrap standard error estimates can suffer from poor coverage \citep{davison1997bootstrap}.  For complex survey design data, valid application of the bootstrap is less straightforward. \citet{baldoni2021use} introduced a multiple imputation (MI)-based procedure for variance estimation when applying regression calibration to complex survey designs, but this method occasionally yielded overly wide confidence intervals. The sandwich variance estimator has also been extended for complex survey designs \citep{binder1983variances,lumley2017fitting}. Likely due to the lack of software for specialized two-stage settings, sandwich variance estimation has rarely been applied in the setting of regression calibration and simple random sampling, and we are not aware of any example in existing literature where this variance approach has been used for regression calibration in a complex survey design. 

In this paper, we develop a simple method to calculate the sandwich variance estimator in two-stage regression settings. This work is motivated by applications in two US cohort studies: the Women's Health Initiative (WHI) and the Hispanic Community Health Study/Study of Latinos (HCHS/SOL), which we describe in more detail in the next section. We then introduce the two-stage model setting and review the stacked estimating approach of \citet{boos2013essential}, which provides a sandwich variance estimate for the outcome model regression parameters.  We develop a procedure that uses quantities returned from the fitted regression and other standard functions in R Software to compute the sandwich estimator, which can be implemented using our package \texttt{sandwich2stage}. We use a simulation study to assess how the sandwich compares to its competing variance estimators, considering scenarios in regression calibration where the bootstrap has coverage problems. Finally, we report the analyses of our data examples and conclude by summarizing our findings.

\section{Motivating Data Examples}

The Women's Health Initiative (WHI), a collection of studies launched in 1993, investigated the major causes of morbidity and mortality in US post-menopausal women aged 50-79 \citep{study1998design}. It is of interest to assess the association of incident diabetes with dietary energy, protein, and protein density consumption, where these exposures are self-reported and error-prone. The WHI also included the Nutritional Biomarker Study, in which objective recovery biomarkers for energy and protein intake were recorded on a subset of participants \citet{neuhouser2008use}. \citet{tinker2011biomarker} used regression calibration to correct for error in self-reported dietary energy and protein and then evaluated the association between the calibrated dietary variables and incident diabetes using a Cox proportional hazards model. Variance estimates for the outcome model parameters were obtained using the bootstrap. We reanalyze data from \citet{tinker2011biomarker} and compare the bootstrap and sandwich variance estimates.

For the complex survey setting, we consider an example from the Hispanic Community Health Study/Study of Latinos (HCHS/SOL), an ongoing, multicenter community-based cohort study of Hispanics/Latino adults aged 18-74 years recruited from randomly selected households at 4 US field centers (Chicago, IL; Miami, FL; Bronx, NY; San Diego, CA). The HCHS/SOL cohort was recruited using a multi-stage, probability-based sampling design. A random subset of HCHS/SOL participants was enrolled in the Study of Latinos: Nutrition and Physical Activity Assessment Study (SOLNAS), and objective recovery biomarkers were collected for several dietary components \citep{mossavar2015applying}. \citet{baldoni2021use} used regression calibration to correct for error in self-reported dietary potassium in the HCHS/SOL and assessed the cross-sectional association between calibrated potassium intake and baseline hypertension-related outcome variables. We reanalyze this data to compare the variance estimates from MI and the sandwich.

\section{Methods}

This section begins by introducing the two-stage model setup. For ease of presentation, we present two-stage regression in the context of regression calibration, where at stage 1, nuisance parameters are estimated from a model used to adjust (calibrate) the observed exposure, and in stage 2, an outcome model is fit using the plug-in estimator from stage 1.  We assume the standard setup for regression calibration, where the stage 1 model may be fit only on a subset of individuals but the stage 2  model is fit on the whole study cohort. We describe the proposed sandwich estimator and the established competing variance estimators.

\subsection{Notation and two-stage model setup}

We consider the two-stage model setting, where a $j \times 1 $ vector $\boldsymbol{\alpha}$ and the $k \times 1$ vector $\boldsymbol{\beta}$ are estimated at stage 1 and stage 2, respectively. Consider a study cohort of $N$ individuals, either from a SRS or complex survey. For $i=1,\ldots,N$, let $X_i^*$ be an observed, error-prone covariate, which we assume is linearly related with the true, unobserved exposure $X_i$ and other error-free covariates $Z_i$ such that

\begin{equation}\label{regcalmodelpaper3}
    X_i=\alpha_{0}+\alpha_{1}X^{*}_i+\alpha_{2}'Z_i+U_i,
\end{equation}

\noindent where $U_i$ is a random error term, independent of all variables, with mean 0 and variance $\sigma_{U}^2$, and $\alpha=(\alpha_0,\alpha_1,\alpha_2)$. Suppose there is a sub-study of size $n$ where $X_i^{**}$ is observed that follows the classical measurement error model, i.e.

\begin{equation}\label{classicalmepaper3}
X_i^{**}=X_i+\epsilon_i,
\end{equation}

\noindent where $\epsilon_i$ is a random error, independent of all variables, with mean 0 and variance $\sigma_{\epsilon}^2$. Define $V_i$ as the indicator individual $i$ is in the sub-study used to fit the stage 1 model.

\subsection{Stage 1 Model}

Denote the stage 1 model of interest by $f(x^{**}|x^*,z; \boldsymbol{\alpha})$, the conditional probability density function of $X^{**}_i$ given $X^*_i$ and $Z_i$. Data for the stage 1 model consist of $\{X_i^{**},X_i^*,Z_i\}$ for the $n$ individuals in the sub-study. Let $l_{1}(\boldsymbol{\boldsymbol{\alpha}})$ be the corresponding log-likelihood for the stage 1 model. To apply regression calibration, one builds a calibration model to estimate the average true exposure given the observed covariates, namely $\hat{X}_i(\boldsymbol{\alpha})=E(X_i|X^*_i,Z_i;\boldsymbol{\alpha})$, which is the stage 1 model. For ease of notation, we often suppress $\boldsymbol{\alpha}$ and use $\hat{X}_i$ to denote both the model $\hat{X}_i(\boldsymbol{\alpha})$ and the fitted $\hat{X}_i(\hat{\boldsymbol{\alpha}})$, which uses the plug-in estimator $\hat{\boldsymbol{\alpha}}$. The calibration model parameters may be estimated if $X_i$ or $X_i^{**}$, a measure containing independent classical error, is observed on a subset.

\subsection{Stage 2 Model}\label{stage2}

Suppose we are interested in the relationship between outcome $Y_i$ and the covariates $(X_i,Z_i)$. 
Let $g(y|x,z; \boldsymbol{\alpha}, \boldsymbol{\beta})$ be the stage 2 model of interest, which has corresponding log-likelihood function $l_{2}(\boldsymbol{\beta})$. The stage 2 model is fit using the $N$ observations $\{Y_i, X_i,Z_i\}$ from the main study. When the exposure of interest is unobserved, one can substitute the estimated $\hat{X}_i$ in for $X_i$ in the outcome model to obtain an estimate of the unknown regression parameter vector, $\boldsymbol{\beta}$ \citep{prentice1982covariate,keogh20}. This is accomplished by using the estimated plug-in parameters, $\hat{\boldsymbol{\alpha}}$, from stage 1 to estimate $\hat{X}_i$.

\subsection{Variance Estimation}

We now describe the different variance estimators that incorporate the uncertainty added by estimating the nuisance parameters from stage 1. We present the formulation of the sandwich variance estimator \citep{boos2013essential} for the SRS and complex survey design settings and provide a convenient estimation method in R software. We also review the bootstrap variance estimator \citep{efron1979bootstrap} and the MI procedure proposed by \citet{baldoni2021use}.

\subsubsection{Sandwich Variance Estimator}\label{SRS}

\noindent \textit{Case 1: Simple Random Sample}\\~\

The sandwich variance estimator is obtained by stacking the calibration and outcome model estimating equations \citep{boos2013essential}. We are interested in the  ``stacked'' parameter vector, $\theta=(\boldsymbol{\alpha},\boldsymbol{\beta})$, which includes the parameters from the stage 1 and stage 2 models. We introduce these methods for a subset of outcome models, e.g. the familiar generalized linear model (GLM) and Cox proportional hazards model, but our technique  works more generally for any pair of stage 1 and stage 2 models for which a sandwich variance estimator exists.  We outline sufficient assumptions for sandwich variance estimation in Section S1 in the Supplementary Materials. 

Define $U_i(\theta)$  as the $j+k$-dimensional vector of stacked estimating equations formed for $\theta$, which can be broken down into the estimating equations for the stage 1 model, $U_{i1}(\theta)$, and the stage 2 model, $U_{i2}(\theta)$.  For maximum likelihood estimation, $U_{i1}(\theta)$ and $U_{i2}(\theta)$ are the score functions, or the vector of first derivatives of the log-likelihood functions $l_{i1}(\boldsymbol{\alpha})$ and $l_{i2}(\boldsymbol{\beta})$, respectively, with respect to the parameters being estimated. Note, $l_{i1} = 0$ for those not in the stage 1 model. We now write $U_i(\theta)$ as:

\begin{equation}\label{stackedEstEq}
    U_i(\theta)= \begin{bmatrix}[1.5]  U_{i1}(\theta) \\   U_{i2}(\theta)  \end{bmatrix}= \begin{bmatrix}[1.5] \frac{\partial l_{i1}(\boldsymbol{\alpha})}{\partial \boldsymbol{\alpha}} \\   \frac{\partial l_{i2}(\boldsymbol{\beta})}{\partial \boldsymbol{\beta}} \end{bmatrix}.
\end{equation}

Estimates of our vector of unknown parameters, $\hat{\theta}$, can be found by solving the equations $\sum_{i=1}^NU_i(\theta)=0$. Following \citet{boos2013essential}, a sandwich estimator for the variance of $\hat{\theta}$ takes the form:

\begin{equation}\label{sandwich}
    V(\hat{\theta})=A(\hat{\theta})^{-1} B(\hat{\theta}) \left[A(\hat{\theta})^{-1}\right]^T/N,
\end{equation}

\noindent where

\begin{equation}
    A(\hat{\theta})=\frac{1}{N}\sum_{i=1}^N \left.\frac{\partial U_i(\theta)}{\partial \theta}\right|_{\theta=\hat{\theta}},
\end{equation}

\noindent and 

\begin{equation}
    B(\hat{\theta})=\frac{1}{N}\sum_{i=1}^N U_i(\hat{\theta})U_i(\hat{\theta})^T.
\end{equation}

The $[j+k] \times [j+k]$ matrix  $A(\hat{\theta})$ has the following form:

\begin{equation}\label{myAMAT}
A(\hat{\theta})=\frac{1}{N}\sum_{i=1}^N\begin{bmatrix}[1.5]
\frac{\partial U_{i1}(\theta)}{\partial \boldsymbol{\alpha}}|_{\theta=\hat{\theta}} & 
\frac{\partial U_{i1}(\theta)}{\partial \boldsymbol{\beta} }|_{\theta=\hat{\theta}} \\
\frac{\partial U_{i2}(\theta)}{\partial  \boldsymbol{\alpha}}|_{\theta=\hat{\theta}} &
\frac{\partial U_{i2}(\theta)}{\partial  \boldsymbol{\beta}  } |_{\theta=\hat{\theta}}

\end{bmatrix}=\frac{1}{N}\sum_{i=1}^N\begin{bmatrix}[1.5]
\frac{\partial U_{i1}(\theta)}{\partial \boldsymbol{\alpha}}|_{\theta=\hat{\theta}} & 
0 \\
\frac{\partial U_{i2}(\theta)}{\partial  \boldsymbol{\alpha}}|_{\theta=\hat{\theta}} &
\frac{\partial U_{i2}(\theta)}{\partial  \boldsymbol{\beta}  } |_{\theta=\hat{\theta}}
\end{bmatrix}.
\end{equation}

In the case of maximum likelihood estimation, the $[j \times j]$  upper-left quadrant, $\frac{\partial U_{i1}(\theta)}{\partial \boldsymbol{\alpha}}|_{\theta=\hat{\theta}}$, and $[k \times k]$ bottom-right quadrant, $\frac{\partial U_{i2}(\theta)}{\partial \boldsymbol{\beta}}|_{\theta=\hat{\theta}}$, are the Hessian matrices for the stage 1 and stage 2 models, respectively.  All elements of the $[j \times k]$ upper-right quadrant, $\frac{\partial U_{i1}(\theta)}{\partial \boldsymbol{\beta} }|_{\theta=\hat{\theta}}$, are 0 because the stage 2 parameters are not in the estimating equation for the stage 1 model. The $[k \times j]$ elements of the bottom-left quadrant, $\frac{\partial U_{i2}(\theta)}{\partial  \boldsymbol{\alpha}}|_{\theta=\hat{\theta}}$, are non-zero since the estimated exposure $\hat{X}_i$ in the stage 2 model relies on the $\boldsymbol{\alpha}$ parameters from stage 1. These derivatives can be computed directly if a closed-form solution exists or using numerical derivatives. The $[j+k] \times [j+k]$ matrix  $B(\hat{\theta})$ has the following general form:

\begin{equation}
    B(\hat{\theta}) = \frac{1}{N}\sum_{i=1}^N    \begin{bmatrix}[1.5] U_{i1}(\hat{\theta}) \\  U_{i2}(\hat{\theta})  \end{bmatrix} \begin{bmatrix}[1.5]U_{i1}^T(\hat{\theta}) &  U_{i2}^T(\hat{\theta})  \end{bmatrix}= \frac{1}{N}\sum_{i=1}^N    \begin{bmatrix}[1.5] U_{i1}(\hat{\theta})U_{i1}^T (\hat{\theta})  & U_{i1}(\hat{\theta})U_{i2}^T (\hat{\theta}) \\  U_{i2}(\hat{\theta})U_{i1}^T (\hat{\theta})& U_{i2}(\hat{\theta}) U_{i2}^T (\hat{\theta})  \end{bmatrix}.
\end{equation}

In R, this sandwich variance estimate can be obtained by (1) directly computing the matrices in equation \ref{sandwich}, or (2) taking advantage of convenient functions in the survey package and readily available quantities from the fitted stage 1 and stage 2 models \citep{lumley2011complex}.  For the latter, we reformulate the sandwich variance  in terms of the influence functions $A(\hat{\theta})^{-1}\tilde{U}(\hat{\theta})$, where $\tilde{U}(\hat{\theta})$ is the matrix of transposed estimating equation contributions, i.e.:

\begin{eqnarray}\label{Ubar}
   \tilde{U}(\hat{\theta}) = \begin{bmatrix}  U_{11}^T(\hat{\theta}) & U_{12}^T(\hat{\theta})  \\
    U_{21}^T(\hat{\theta}) &  U_{22}^T(\hat{\theta})  \\
    ... & ...  \\
    U_{(N-1)1}^T(\hat{\theta}) &  U_{(N-1)2}^T(\hat{\theta})  \\ U_{N1}^T(\hat{\theta}) & U_{N2}^T(\hat{\theta})  \\
    \end{bmatrix} 
    \end{eqnarray}.

In Figure \ref{fig:code1}, we provide sample R code that computes $V(\hat{\theta)}$ using the influence function approach. Our R function assumes a linear model for stage 1, fit using \texttt{svyglm} with a gaussian response. We assume the stage 2 model is a generalized linear or Cox model fit using the \texttt{svyglm} or \texttt{svycoxph} functions. In Figure \ref{fig:models}, we create the SRS survey design and provide sample model fitting statements for stage 1 and stage 2. In Figure \ref{fig:Amat}, we show how to obtain the pieces of the matrix, $A(\hat{\theta})$. The diagonal components of $A(\hat{\theta})$ are extracted from the stage 1 and 2 model fits; for the bottom-left, we developed the R function \texttt{stage2.alphas.estfuns()}, which computes the numerical derivatives for the stage 2 estimating equation with respect to $\boldsymbol{\alpha}$. Next in Figure \ref{fig:Umat}, we create the matrix $\tilde{U}(\hat{\theta})$ by stacking the estimating equation contributions for the stage 1 and stage 2 models, extracted from the model objects using \texttt{estfun()}. All stage 1 estimating equation contributions for those not in the calibration subset are equal to 0. Finally, in Figure \ref{fig:svyinflsand} we compute the influence functions by multiplying the $[N \times (j+k)]$ matrix $\tilde{U}(\hat{\theta})$ by the $[j+k] \times [j+k]$ matrix $\left[A(\hat{\theta})^{-1}\right]^T$ and dividing by the sample size, $N$, which are given to \texttt{vcov(svytotal())} along with the study design to compute an estimate of $V(\hat{\theta})$. Alternatively, one can use our package \texttt{sandwich2stage}, which accomplishes each of the aforementioned steps to calculate the sandwich variance. Our function requires the fitted stage 1 and 2 model objects, the names of the variables $X^*_i$ and $\hat{X}_i$ in the data, and the names of the subject ID variables for stage 1 and 2, as illustrated in Figure \ref{fig:sandpackage}.\\~\
     
\noindent \textit{Case 2: Complex Survey Design}\\~\

Our two-stage sandwich variance estimator can easily be adapted for a complex survey design. We consider designs in which the validation sub-study is nested in the full study cohort, but this approach may be easily extended to other designs. Define $\pi_i$ as the probability that subject $i$ is included in the sample, which is known from the survey design. A participant sampled with probability $\pi_i$ is assumed to represent $1/\pi_i$ participants in the total population, which becomes the sampling weight reflecting unequal probability of selection into the sample \citep{lumley2011complex}. Consider $\widecheck{U_{i}}(\theta)=\frac{1}{\pi}U_{i}(\theta)$, the vector of stacked estimating equations formed for $\theta$ in a probability-based sampling design. As before, $\widecheck{U_{i}}(\theta)$ can be divided into  $\widecheck{U}_{i1}(\theta)$ and  $\widecheck{U}_{i2}(\theta)$. Define $\widecheck{l}_{1}(\boldsymbol{\alpha})=\frac{1}{\pi}l_{1}(\boldsymbol{\alpha})$ and  $\widecheck{l}_{2}(\boldsymbol{\beta})=\frac{1}{\pi}l_{2}(\boldsymbol{\beta})$ as the weighted log-likelihood for the stage 1 and stage 2 models, respectively. The $j+k$-dimensional vector of stacked estimating equations is then:

\begin{equation}
   \widecheck{U}_{i}(\theta)= \begin{bmatrix}[1.5]\widecheck{U}_{i1}(\theta) \\  \widecheck{U}_{i2}(\theta)  \end{bmatrix}= \begin{bmatrix}[1.5] \frac{\partial \widecheck{l}_{1}(\boldsymbol{\alpha})}{\partial \boldsymbol{\alpha}} \\   \frac{\partial \widecheck{l}_{2}(\boldsymbol{\beta})}{\partial \boldsymbol{\beta}} \end{bmatrix}.
\end{equation}

To obtain estimates of the unknown, design-based parameters, $\hat{\theta}$, one can solve $\sum_{i=1}^N\widecheck{U}_i(\theta)=0$. \citet{binder1983variances} applied a standard delta method argument to provide a sandwich form for the estimated design-based variance, $V_{\pi}(\hat{\theta})=A_{\pi}(\hat{\theta})^{-1} B_{\pi}(\hat{\theta}) \left[A_{\pi}(\hat{\theta})^{-1}\right]^T/N$. Following \citet{lumley2017fitting}, 

\begin{equation}
    A_{\pi}(\hat{\theta})=\frac{1}{N}\sum_{i=1}^N  \left.\frac{\partial \widecheck{U}_i(\theta)}{\partial \theta}\right|_{\theta=\hat{\theta}}
\end{equation}

\noindent and 

\begin{equation}
    B_{\pi}(\hat{\theta})= \widehat{\textrm{var}}_{\pi} \left[\frac{1}{N}\sum_{i=1}^N  \widecheck{U}_i(\hat{\theta}) \right].
\end{equation}

In Figure \ref{fig:code2}, we provide sample code that computes the sandwich for a complex survey design with stratification, clustering, and unequal probability weighting, again using our convenient influence function approach. We first create a hypothetical survey design object, then fit the stage 1 and stage 2 models. Since \texttt{svytotal()} adds the weights, one must provide the unweighted matrix, $\tilde{U}(\hat{\theta})$, which can be accomplished by dividing the estimating functions by the weights. 

For the stacked estimating equation, it is important to consider the strata arising from (1) original sampling procedure of the survey design and (2) membership to the calibration subset, if this wasn't one of the original strata by design. In this setting, one can augment the original design strata by cross-classifying the design strata with the subset indicator.

\subsubsection{Bootstrap}\label{SRSboot}

Bootstrap variance estimation is often implemented in two-stage regression due to its simplicity. One step that is often overlooked is ensuring that bootstrap sampling is stratified on subset membership status. For those in the calibration subset, $\{X_i^*,X_i^{**},Z_i,Y_i\}$ is resampled. Otherwise, $\{X_i^*,Z_i,Y_i\}$ is resampled. The stage 1 model is then fit to the bootstrap sample to obtain a new exposure variable estimate, which is subsequently included in the stage 2 model fit to the bootstrap sample. Section S2 in the Supplementary Materials includes steps for computing bootstrap standard errors using a stratified bootstrap procedure.

\subsubsection{Multiple Imputation}\label{surveyMI}

\citet{baldoni2021use} proposed a variance estimation approach that multiply imputes the expected value of the latent exposure variable for all individuals in the main study by repeatedly sampling the stage 1 model coefficients required to estimate $\hat{X}_i$. New calibration coefficients can be sampled using either their estimated asymptotic parametric distribution or bootstrap resampling.  At each imputation step, the outcome model is re-fit using the newly calibrated values. Details on the resampling-based MI approach are provided in section S3 of the Supplementary Materials.

\section{Simulation Study}
\subsection{Setup}
We use a simulation study to compare the sandwich to competing variance estimators for the SRS and complex survey settings. We let the stage 1 model ($n=450$) and stage 2 model ($N=1000$ or $N=10,000$) be linear and logistic regression  models, respectively. We generate two covariates, $X_i$ and $Z_i$, from a multivariate normal distribution with low or high correlation. We simulate an error-prone covariate $X_i^*$ and vary the error variance $\sigma^2$ between 0, 0.25, 0.50, and 1.00 to represent zero, low, moderate, and high measurement error, respectively. We also conduct complex survey based simulations using code provided by \citet{baldoni2021use} for simulating a superpopulation and drawing samples using a stratified complex survey sampling scheme. Full simulation details are provided in Supplementary Materials Section S4. In all tables, we present median percent ($\%$) biases, medians of the estimated standard errors (ASE), empirical median absolute deviations of the estimated regression coefficients (MAD), and 95\% coverage probabilities (CP).

\subsection{Results}

In Table \ref{table1simspaper3}, we present results for the stage 2 model fit to the true exposure $X_i$, the error-prone exposure $X_i^*$, and the calibrated exposure $\hat{X}_i$ for data simulated from a SRS. For the regression calibration approaches, we compute model-based (naive), sandwich, and bootstrap standard errors. Applying regression calibration reduces the absolute median bias to under 6\% in all settings. For the larger samples ($N=10,000$), using naive standard errors with regression calibration results in CP as low as 84\%. Applying sandwich or bootstrap techniques gives ASEs that more closely resemble the empirical MAD values and CP closer to 95\%.  In the high correlation and error settings, CP calculated by the bootstrap is too high ($>97\%$), while the sandwich maintains the nominal coverage level (CP$=95\%$). Similar patterns were observed for simulations using a Cox proportional hazards model in stage 2, where the sandwich performed well and occasionally outperformed the bootstrap (Supplementary Table S1). In Table S2 in the Supplementary Materials, we compute CP from confidence intervals constructed using the standard Wald procedure with bootstrap standard errors, bootstrap distribution percentiles, and the bias-corrected and accelerated (BCA) bootstrap approach. In high correlation and error settings where intervals constructed using bootstrap standard errors resulted in CP that was too large, the percentile and BCA bootstrap also present an opportunity to improve. 

In Table \ref{table2simspaper3}, we present results for data simulated from a complex survey. Once again, the naive model-based standard errors are frequently too small, resulting in $CP < 95\%$. Standard error estimates obtained from the sandwich are generally better-behaved than those estimated from the resampling-based MI approach of \citet{baldoni2021use}. The standard errors estimated from  MI are oftentimes too large, resulting in CP as high as 99\% for the high measurement error and correlation case when $N=1000$. These same issues of over-coverage were observed by \citet{baldoni2021use}, which the authors attempted to mitigate by using robust estimators for the mean and standard deviation in their calculation of the adjusted variance. While these robust estimators did improve the estimated variances, they did not completely address over-coverage issues. 

\section{Reanalysis of WHI and HCHS/SOL Data}

\subsection{WHI Data Example}

To assess the association between energy, protein, and protein density with the risk of diabetes in the WHI, we begin by developing our stage 1 model using data from the Nutritional Biomarker Study. The stage 1 model was fit to $n=356$ sub-study participants as a linear regression of the biomarker value $(X^{**})$ on the corresponding self-reported value $(X^{*})$ and covariates $(Z)$. The stage 2 model included $N=77,805$ eligible women and was fit as a Cox proportional hazards model. Following \citet{tinker2011biomarker}, we fit the stage 2 model with and without BMI and report Wald confidence intervals based on the bootstrap SE. Details on our models and analytic cohort are provided in Section S5 of the Supplementary Materials. 

Incident diabetes was reported in 4278 (5.5\%) of the $77,805$ participants in the analytic cohort. Table \ref{WHIdataanalysisall} presents hazard ratio (HR) estimates and 95\% confidence intervals (CI) for incident diabetes for a 20\% increase in consumption of energy, protein, and protein density, as well as estimated $\beta$ coefficients and standard errors (SEs). In the BMI-adjusted analysis, the HR for a 20\% increase in calibrated energy intake was 1.54. The naive SE estimate on the log scale, 1.37, is over 40\% smaller than the sandwich SE estimates, 2.34, resulting in a 95\% CI for the HR for a 20\% increase in energy intake with the naive SE of (0.94, 2.52), compared to (0.68, 3.51) using the sandwich. The bootstrap SE estimate for this model is 36.97, corresponding to a 95\% CI of (0, 842640) for a 20\% increase in energy consumption. We note that 47 (9.4\%) of our 500 bootstrap replicates resulted in  $|\hat{\beta}|>10$ for log-energy, which is impractical from an epidemiologic perspective. As an alternative, we present 95\% CIs constructed using the percentile bootstrap interval, which is (0.04, 37.17) for a 20\% increase in energy in the BMI-adjusted model. In this example, the sandwich offers a more believable standard error estimate. 

Differences between the sandwich and bootstrap were not as extreme for the non-BMI adjusted energy analysis. Similar patterns were observed for protein, where the HR for a 20\% increase in calibrated intake adjusted for BMI was 1.23, with 95\% CIs of (1.08, 1.40) for the sandwich, (1.03, 1.46) for the normal bootstrap, and (1.11, 1.52) for the percentile bootstrap. In the BMI-adjusted models with log-protein density, the HR (95\% CI) estimated by regression calibration with the naive SE is 1.64 (1.42, 1.88) compared to CIs of (1.14, 2.34), (0.83, 3.24), (1.31, 3.93) estimated by the sandwich, normal bootstrap, and percentile bootstrap, respectively. In this instance, using the bootstrap SE (1.91) resulted in a loss of statistical significance, which did not occur with the sandwich SE (1.00) or the bootstrap percentile interval. We note instability of the bootstrap SE was not observed in the original analysis by \citet{tinker2011biomarker}. Differences between our results and those reported by \citet{tinker2011biomarker} are discussed in the Supplementary Materials.

\subsection{HCHS/SOL Data Example}

In the HCHS/SOL study, we fit the stage 1 model using data from $n=310$ SOLNAS subset members by performing a linear regression of the log-potassium biomarker ($X^{**}$) on the log self-reported 24-hour recall ($X^*$) and other covariates ($Z$). We consider hypertension status and systolic blood pressure, which are studied using logistic and linear stage 2 models, respectively. For each model, the outcome was regressed on the calibrated dietary exposure while adjusting for confounders. All stage 2 outcome models were fit to a subset of $N=8,176$ participants from the original cohort and accounted for the HCHS/SOL survey design. Details on our models and data are included in Section S6 of the Supplementary Materials.

Table \ref{HCHSdataanalysisall} presents results from our re-analysis of the HCHS/SOL data. The estimated odds ratio (OR) of hypertension for a 20\% increase in calibrated potassium (95\% CI with naive SE) is 0.81 (0.63, 1.03). The naive SE estimate on the log scale of 0.68 is 15\% and 44\% smaller than the SEs estimated by the sandwich (0.80) and MI (1.22), respectively. Similar patterns were observed for the linear regression of systolic blood pressure on calibrated log-potassium, where the estimated coefficient for a 20\% increase was -0.58, with 95\% CIs of (-1.26, 0.10) with the naive SE, (-1.37, 0.21) for the sandwich, and (-1.78, 0.62) for MI. The SE estimated by MI, 3.35, was more inflated compared to the sandwich SE, 2.21, for this particular model.

\section{Discussion}

In this manuscript, we increase the practicality of the sandwich for standard error estimation in two-stage regression settings and compare it to competing methods. We use two data examples to illustrate the straightforward use of the sandwich for data from a SRS and complex survey design. We developed R code, which we hope makes computation of the sandwich more accessible and convenient. For obtaining the sandwich in R, one can modify the code provided throughout this paper or use functions from our package on GitHub at \url{https://github.com/lboe23/sandwich2stage}, which directly computes sandwich variance estimates for the two-stage setting of regression calibration.

Our numerical study indicated that when the sub-study is a large percentage of the main study, ignoring the uncertainty in the estimation of the stage 1 model parameters does not have a large impact, resulting in well-behaved ``naive" standard errors. This point is also discussed by \citet{keogh20}. We do encourage readers to adjust the naive model standard errors in all two-stage regression model settings, however, as naive standard errors will be too small in many instances.

In addition to the added computational burden, bootstrap confidence interval estimation can suffer from poor coverage if appropriate bias-adjustment procedures are not applied to departures from normality in bootstrap estimates \citep{efron1987better}. Despite the well-known bias and coverage problems of the standard normal Wald bootstrap interval, the BCA procedure is rarely applied in practice due to complexity. In our data example from the WHI, the bootstrap SE in the BMI-adjusted analysis for calibrated energy was quite unstable, resulting in extremely inflated 95\% CIs compared to the sandwich. Nonetheless, the sandwich estimator also has some limitations, and occasionally showed coverage problems in settings with substantial measurement error and highly correlated covariates. The BCA bootstrap procedure may be the optimal approach for constructing 95\% CIs. However, this technique can be computationally prohibitive in large cohorts like the WHI. In our complex survey simulations,  confidence intervals calculated using MI resulted in overcoverage, likely from instability caused by multicollinearity in the simulated data sets \citep{baldoni2021use}. We also observed inflated MI-based standard errors compared to the sandwich in our HCHS/SOL data example. Instability in standard error estimates from the resampling-based bootstrap and MI procedures can be easily avoided if a more stable estimator like the sandwich is used for variance estimation. 

This paper focused on regression calibration as a motivating example, but the issues discussed apply more broadly to any two-stage regression setting where plug-in estimates are obtained in stage 1 and used in the stage 2 model. Generally speaking, in two-stage regression settings where model-based standard errors are too small and standard errors obtained from resampling based approaches can be unstable, the sandwich variance approach presents a well-behaved, less computationally-intensive alternative that is straightforward to implement. The sandwich may also be extended to multi-stage regression models. One related example in the WHI is a sandwich estimator used in regression calibration dietary applications when a third component to the stacked estimating equations is added for biomarker development \citep{prentice2021biomarker, prentice2022biomarkers}. Future work will extend the software for two-stage regression to these multi-stage model settings.

\newpage

\section*{Acknowledgements}
This work was supported in part by NIH grant R01-AI131771. The authors would like to thank the investigators of the Women’s Health Initiative (WHI) and the Hispanic Community Health Study/Study of Latinos (HCHS/SOL) for the use of their data. A short list of WHI investigators can be found here: \url{https://www.whi.org/doc/WHI-Investigator-Short-List.pdf}. 
A list of HCHS/SOL investigators, managers and coordinators by field center can be found here: \url{https://sites.cscc.unc.edu/hchs/Acknowledgement}. {\it Conflict of Interest}: None declared.

\section*{Data Availability Statement}
The data sets used in this paper were obtained through submission and approval of manuscript proposals to (1) the Women's Health Initiative Publications and Presentations Committee, as described on the WHI website \citep{WHIdatacite} and (2) the Hispanic Community Health Study/Study of Latinos Publications Committee, as described on the HCHS/SOL website \citep{SOLdatacite}. For more details on each of these processes, see  \url{https://www.whi.org/page/propose-a-paper} and \url{https://sites.cscc.unc.edu/hchs/publications-pub}.

\newpage

\begin{figure}[hbtp]
         \caption{Code for obtaining the sandwich variance for a simple random sample in the R software}
        \label{fig:code1}
        \begin{subfigure}[b]{1\textwidth}
       \caption{Code for fitting stage 1 and stage 2 models}
\label{fig:models}
\begin{lstlisting}
   sampdesign <- svydesign(id=~1, data=mydata)
   stage1.model<-svyglm(xstarstar~xstar+z,design=sampdesign,
        family=gaussian(),subset=V==1)
   alphas.stage1<-coef(stage1.model)
   sampdesign <- update(sampdesign,xhat =predict(stage1.model,
        newdata=sampdesign$variables) )
   stage2.model<- svycoxph(Surv(Time, delta) ~ xhat+z, 
        design=sampdesign)
    \end{lstlisting} 
\end{subfigure}
    \begin{subfigure}[b]{1\textwidth}
       \caption{Code for obtaining $A(\hat{\theta})$}
\label{fig:Amat}
\begin{lstlisting}
    A.upperleft<- -solve(stage1.model$naive.cov)/N
    A.bottomright<- -solve(stage2.model$naive.cov)/N
    A.upperright<- matrix(0,nrow=j,ncol=k)
    A.bottomleft<- stage2.alphas.estfuns(alphas.stage1)
    A<-rbind(cbind(A.upperleft,A.upperright),
        cbind(A.bottomleft,A.bottomright))
    A.inv<-solve(A)
    \end{lstlisting} 
\end{subfigure}
\begin{subfigure}[b]{1\textwidth}
    \caption{Code for obtaining $\tilde{U}(\hat{\theta})$}
\label{fig:Umat}
\begin{lstlisting}
    estfun.stage1<-matrix(0,nrow=N,ncol=j)
    is.calibration <- !is.na(xstarstar)
    estfun.stage1[is.calibration,] <- estfun(stage1.model)
    estfun.stage2<-as.matrix(estfun(stage2.model))
    estfun.all<-cbind(estfun.stage1,estfun.stage2)
    \end{lstlisting}
     \end{subfigure}
     \begin{subfigure}[b]{1\textwidth}
       \caption{Code for obtaining $V(\hat{\theta})$}
       \label{fig:svyinflsand}
      \begin{lstlisting}
    infl<- as.matrix(estfun.all)%*%t(A.inv)/N
    sandwichvar<-vcov(svytotal(infl, sampdesign))
\end{lstlisting}
     \end{subfigure}
     \begin{subfigure}[b]{1\textwidth}
       \caption{Code for accomplishing steps (b)-(d) using the package \texttt{sandwich2stage}} 
\label{fig:sandpackage}
\begin{lstlisting}
    sandwich.object<-sandwich2stage(stage1.model,stage2.model,
        xstar="xstar",xhat="xhat",
        Stage1ID="ID",Stage2ID="ID")
    sandwichvar<-vcov(sandwich.object)
    \end{lstlisting} 
\end{subfigure}
     \end{figure}

\begin{figure}[hbtp]
         \caption{Code for obtaining the sandwich variance for a complex survey design in the R software}
        \label{fig:code2}
        \begin{subfigure}[b]{1\textwidth}
       \caption{Code for fitting stage 1 and stage 2 models}
\label{fig:modelssvy}
\begin{lstlisting}
   sampdesign <- svydesign(id=~PSUid, strata=~strat, 
                weights=~myweights, data=mydata)  
   stage1.model<-svyglm(xstarstar~xstar+z,design=sampdesign,
        family=gaussian(),subset=V==1)
   alphas.stage1<-coef(stage1.model)
   sampdesign <- update(sampdesign,xhat =predict(stage1.model,
        newdata=sampdesign$variables) )
   stage2.model<- svyglm(y ~ xhat+z ,
        design=sampdesign,family=binomial())
    \end{lstlisting} 
\end{subfigure}
    \begin{subfigure}[b]{1\textwidth}
       \caption{Code for obtaining $A(\hat{\theta})$}
\label{fig:Amat2}
\begin{lstlisting}
    A.upperleft<- -solve(stage1.model$naive.cov/
                    mean(stage1.model$prior.weights))/N
    A.bottomright<- -solve(stage2.model$naive.cov/
                    mean(stage2.model$prior.weights))/N
    A.upperright<- matrix(0,nrow=j,ncol=k)
    A.bottomleft<- stage2.alphas.estfuns(alphas.stage1)
    A<-rbind(cbind(A.upperleft,A.upperright),
        cbind(A.bottomleft,A.bottomright))
    A.inv<-solve(A)
    \end{lstlisting} 
\end{subfigure}
\begin{subfigure}[b]{1\textwidth}
    \caption{Code for obtaining $\tilde{U}(\hat{\theta})$}
\label{fig:Umat2}
\begin{lstlisting}
    estfun.stage1<-matrix(0,nrow=N,ncol=j)
    is.calibration <- !is.na(xstarstar)
    estfun.stage1[is.calibration,] <- estfun(stage1.model)/
       stage1.model$prior.weights
    estfun.stage2<-estfun(stage2.model)/       
        stage2.model$prior.weights
    estfun.all<-cbind(estfun.stage1,estfun.stage2)
    \end{lstlisting}
     \end{subfigure}
     \begin{subfigure}[b]{1\textwidth}
       \caption{Code for obtaining $V(\hat{\theta})$}
       \label{fig:svyinflsand2}
      \begin{lstlisting}
    infl<- as.matrix(estfun.all)%*%t(A.inv)/N
    sandwichvar<-vcov(svytotal(infl, sampdesign))
\end{lstlisting}
     \end{subfigure}
      \begin{subfigure}[b]{1\textwidth}
       \caption{Code for accomplishing steps (b)-(d) using the package \texttt{sandwich2stage}} 
\label{fig:sandpackagesvy}
\begin{lstlisting}
    sandwich.object<-sandwich2stage(stage1.model,stage2.model,
        xstar="xstar",xhat="xhat",
        Stage1ID="ID",Stage2ID="ID")
    sandwichvar<-vcov(sandwich.object)
    \end{lstlisting} 
\end{subfigure}
     \end{figure}

\begin{table}
\centering
      \begin{threeparttable}[t]
            \caption{The median percent (\%) bias, median standard errors (ASE), empirical median absolute deviation (MAD) and coverage probabilities (CP) for 1000 simulated data sets from a simple random sample for a logistic regression stage 2 model (observed event rate = 0.38) fit to true exposure, naive exposure, and calibrated exposure with naive (model-based) standard errors,  sandwich standard errors, and  bootstrap standard errors. We let $\beta=\log(1.5)$ and vary the correlation between  $X$ and $Z$, the sample size ($N$), and the measurement error variance ($\sigma^2$). Sample size of the calibration subset is $n=450$.}\label{table1simspaper3}
   
        \begin{tabular}{lllcccccccc}

\hline
\multicolumn{3}{l}{} & \multicolumn{4}{c}{Low Correlation}  & \multicolumn{4}{c}{High Correlation}  \\ \cmidrule(r){4-7} \cmidrule{8-11}  $N$ & $\sigma^2$\tnote{(1)} & Method   & \% Bias & MAD & ASE & CP & \% Bias & MAD & ASE & \multicolumn{1}{c}{CP}   \\ 

\hline
 $1000$ & $0.00$ & Truth & \phantom{0}-0.04 & 0.07 & 0.07 & 0.94 & \phantom{-0}0.16 & 0.09 & 0.10 & 0.94 \\

\hline
  & $0.25$  & Naive &  
 -13.78 & 0.12 & 0.11 & 0.91 & -42.84 & 0.12 & 0.12 & 0.68 \\

   &  & RC\tnote{(2)} \hspace*{0.2cm} (Naive SE) &    \phantom{0} -3.76 & 0.13 & 0.12 & 0.94 & \phantom{0} -3.28 & 0.21 & 0.20 & 0.94 \\

   &  & RC\tnote{(2)} \hspace*{0.2cm} (Sandwich) &    \phantom{0}\text{---} &\phantom{0}\text{---} & 0.12 & 0.94 &  \phantom{0}\text{---} & \phantom{0}\text{---} & 0.20 & 0.94 \\

   &  & RC\tnote{(2)} \hspace*{0.2cm} (Bootstrap) &     \phantom{0}\text{---} & \phantom{0}\text{---} &  0.12 & 0.94 &  \phantom{0}\text{---} & \phantom{0}\text{---} & 0.21 & 0.94 \\

   \hline
  
  & $0.50$  & Naive &  -48.54 & 0.08 & 0.08 & 0.36 & -68.36 & 0.09 & 0.09 & 0.13 \\ 
 
   &  & RC\tnote{(2)} \hspace*{0.2cm}  (Naive SE) &   \phantom{0}-4.61 & 0.16 & 0.16 & 0.93 & \phantom{0}-2.90 & 0.28 & 0.27 & 0.94 \\ 

   &  & RC\tnote{(2)} \hspace*{0.2cm} (Sandwich) &    \phantom{0}\text{---} & \phantom{0}\text{---} & 0.16 & 0.93 & \phantom{0}\text{---} & \phantom{0}\text{---} & 0.27 & 0.95 \\

   &  & RC\tnote{(2)} \hspace*{0.2cm} (Bootstrap) &   \phantom{0}\text{---} & \phantom{0}\text{---}  & 0.16 & 0.94 & \phantom{0}\text{---} & \phantom{0}\text{---} & 0.29 & 0.96 \\ 
 
  \hline
  & $1.00$  & Naive &      -71.90 & 0.06 & 0.06 & 0.01 & -83.60 & 0.06 & 0.06 & 0.00 \\

   &  & RC\tnote{(2)} \hspace*{0.2cm}  (Naive SE) &   \phantom{0}-5.31 & 0.22 & 0.21 & 0.94 & \phantom{0}-4.73 & 0.37 & 0.37 & 0.94 \\

   &  & RC\tnote{(2)} \hspace*{0.2cm} (Sandwich) &   \phantom{0}\text{---} &   \phantom{0}\text{---}  & 0.21 & 0.94 & \phantom{0}\text{---} & \phantom{0}\text{---} & 0.37 & 0.95 \\

   &  & RC\tnote{(2)} \hspace*{0.2cm} (Bootstrap) &    \phantom{0}\text{---} &   \phantom{0}\text{---} & 0.22 & 0.96 & \phantom{0}\text{---} & \phantom{0}\text{---} & 0.41 & 0.98 \\

      \hline
 $10000$ & $0.00$ & Truth &   \phantom{-}0.05 & 0.02 & 0.02 & 0.94  &   \phantom{-0}0.05 & 0.03 & 0.03 & 0.94 \\

  &  $0.25$ & Naive &  -12.28 & 0.04 & 0.03 & 0.69 & -41.87 & 0.04 & 0.04 & 0.01 \\ 
 
   &  & RC\tnote{(2)}  \hspace*{0.2cm}   (Naive SE) &    \phantom{0}-1.90 & 0.05 & 0.04 & 0.85 & \phantom{0}-1.15 & 0.08 & 0.06 & 0.89 \\

   &  & RC\tnote{(2)} \hspace*{0.2cm} (Sandwich) &   \phantom{0}\text{---} &  \phantom{0}\text{---} & 0.05 & 0.94 & \phantom{0}\text{---} & \phantom{0}\text{---} & 0.08 & 0.95 \\ 
   &  & RC\tnote{(2)} \hspace*{0.2cm} (Bootstrap) &    \phantom{0}\text{---} &  \phantom{0}\text{---} & 0.05 & 0.94 & \phantom{0}\text{---} & \phantom{0}\text{---} & 0.08 & 0.96 \\

   \hline

  & $0.50$  & Naive &  -47.61 & 0.03 & 0.03 & 0.00 & -67.45 & 0.03 & 0.03 & 0.00 \\

   &  & RC\tnote{(2)} \hspace*{0.2cm} (Naive SE) &   \phantom{0}-2.73 & 0.07 & 0.05 & 0.84 & \phantom{0}-1.68 & 0.11 & 0.08 & 0.89 \\

   &  & RC\tnote{(2)} \hspace*{0.2cm} (Sandwich) &  \phantom{0}\text{---} & \phantom{0}\text{---} & 0.06 & 0.94 & \phantom{0}\text{---} & \phantom{0}\text{---} & 0.10 & 0.95 \\

   &  & RC\tnote{(2)} \hspace*{0.2cm} (Bootstrap) &     \phantom{0}\text{---} &  \phantom{0}\text{---} & 0.07 & 0.95 & \phantom{0}\text{---} & \phantom{0}\text{---} & 0.11 & 0.96 \\ 

     \hline

  & $1.00$  & Naive &     -70.93 & 0.02 & 0.02 & 0.00 & -82.74 & 0.02 & 0.02 & 0.00 \\

   &  & RC\tnote{(2)} \hspace*{0.2cm}  (Naive SE) &    \phantom{0}-2.62 & 0.09 & 0.07 & 0.85 & \phantom{0}-1.61 & 0.15 & 0.12 & 0.88 \\ 

   &  & RC\tnote{(2)} \hspace*{0.2cm} (Sandwich) &    \phantom{0}\text{---} &  \phantom{0}\text{---} & 0.09 & 0.94 & \phantom{0}\text{---} & \phantom{0}\text{---} & 0.14 & 0.95 \\ 
 
   &  & RC\tnote{(2)} \hspace*{0.2cm} (Bootstrap) &     \phantom{0}\text{---} & \phantom{0}\text{---} & 0.09 & 0.95 & \phantom{0}\text{---} & \phantom{0}\text{---} & 0.17 & 0.97 \\ 

  \hline

\end{tabular}
 \begin{tablenotes}
     \item[(1)]  $\sigma^2=$ the variance of the random, normally distributed covariate measurement error
      \item[(2)] RC = Regression calibration
   \end{tablenotes}
    \end{threeparttable}
\end{table}

\begin{table}
\centering
      \begin{threeparttable}[t]
            \caption{The median percent (\%) bias, median standard errors (ASE), empirical median absolute deviation (MAD) and coverage probabilities (CP) for 1000 simulated data sets from a complex survey design for a logistic regression stage 2 model (observed event rate = 0.38) fit to true exposure, naive exposure, and calibrated exposure with naive (model-based) standard errors,  sandwich standard errors, and  MI standard errors. We let $\beta=\log(1.5)$ and vary the correlation between  $X$ and $Z$, the sample size ($N$), and the measurement error variance ($\sigma^2$). Sample size of the calibration subset is $n=450$.}\label{table2simspaper3}
   
        \begin{tabular}{lllcccccccc}
\hline
\multicolumn{3}{l}{} & \multicolumn{4}{c}{Low Correlation}  & \multicolumn{4}{c}{High Correlation}  \\ \cmidrule(r){4-7} \cmidrule{8-11}  $N$ & $\sigma^2$\tnote{(1)} & Method   & \% Bias &  MAD & ASE & CP & \% Bias &  MAD & ASE & \multicolumn{1}{c}{CP}   \\ 

\hline
 $1000$ & $0.00$ & Truth & \phantom{-0}2.53 & 0.11 & 0.11 & 0.95 & \phantom{-0}5.72 & 0.18 & 0.17 & 0.94 \\

  \hline

  & $0.25$  & Naive &       -22.49 & 0.16 & 0.15 & 0.91 & -57.99 & 0.17 & 0.17 & 0.70 \\ 
  
  &  & RC\tnote{(2)} \hspace*{0.2cm} (Naive SE) & \phantom{0-}2.92 & 0.21 & 0.21 & 0.94 & \phantom{0-}6.88 & 0.46 & 0.45 & 0.94 \\

    &  & RC\tnote{(2)} \hspace*{0.2cm} (Sandwich) &   \phantom{0}\text{---} & \phantom{0}\text{---} &   0.21 & 0.95  &  \phantom{0}\text{---} & \phantom{0}\text{---} & 0.46 & 0.96 \\

   &  & RC\tnote{(2)} \hspace*{0.2cm} (MI) &  \phantom{0}\text{---} & \phantom{0}\text{---} &  0.22 & 0.96 &  \phantom{0}\text{---} & \phantom{0}\text{---} & 0.50 & 0.98 \\ 
 
   \hline

  &  $0.50$ & Naive &      -54.06 & 0.12 & 0.12 & 0.52 & -76.87 & 0.13 & 0.12 & 0.29 \\

   &  & RC\tnote{(2)} \hspace*{0.2cm} (Naive SE) &  \phantom{0-}2.26 & 0.28 & 0.27 & 0.94 & \phantom{0-}9.24 & 0.64 & 0.61 & 0.94 \\

   &  & RC\tnote{(2)} \hspace*{0.2cm} (Sandwich) &    \phantom{0}\text{---} & \phantom{0}\text{---}  & 0.27 & 0.96 & \phantom{0}\text{---} & \phantom{0}\text{---} & 0.63 & 0.97 \\

   &  & RC\tnote{(2)} \hspace*{0.2cm} (MI) & \phantom{0}\text{---} & \phantom{0}\text{---} &  0.29 & 0.97 & \phantom{0}\text{---} & \phantom{0}\text{---}  & 0.76 & 0.98 \\

  \hline
  & $1.00$  & Naive &    -74.86 & 0.09 & 0.09 & 0.08 & -87.95 & 0.09 & 0.09 & 0.04 \\

   &  & RC\tnote{(2)} \hspace*{0.2cm}  (Naive SE) &   \phantom{0-}3.95 & 0.38 & 0.37 & 0.94 & \phantom{0-}5.29 & 0.87 & 0.84 & 0.93 \\

   &  & RC\tnote{(2)} \hspace*{0.2cm} (Sandwich) &    \phantom{0}\text{---} & \phantom{0}\text{---}  & 0.37 & 0.96 & \phantom{0}\text{---} & \phantom{0}\text{---}  & 0.89 & 0.98 \\

   &  & RC\tnote{(2)} \hspace*{0.2cm} (MI) &     \phantom{0}\text{---} & \phantom{0}\text{---}  & 0.42 & 0.98 & \phantom{0}\text{---} & \phantom{0}\text{---} & 1.41 & 0.99 \\

      \hline
 $10000$ & $0.00$ & Truth &    \phantom{-0}3.12 & 0.04 & 0.04 & 0.94 & \phantom{-0}4.36 & 0.06 & 0.06 & 0.94 \\ 
 
  \hline
  &  $0.25$ & Naive &    -23.11 & 0.05 & 0.05 & 0.54 & -60.35 & 0.05 & 0.05 & 0.01 \\

   &  & RC\tnote{(2)} \hspace*{0.2cm} (Naive SE) &  \phantom{0-}2.74 & 0.08 & 0.07 & 0.87 & \phantom{0-}6.80 & 0.16 & 0.15 & 0.89 \\

   &  & RC\tnote{(2)} \hspace*{0.2cm} (Sandwich) &   \phantom{0}\text{---} & \phantom{0}\text{---} & 0.08 & 0.95 & \phantom{0}\text{---} & \phantom{0}\text{---}  & 0.17 & 0.96 \\

   &  & RC\tnote{(2)} \hspace*{0.2cm} (MI) &     \phantom{0}\text{---} & \phantom{0}\text{---} &  0.08 & 0.96 & \phantom{0}\text{---} & \phantom{0}\text{---}  & 0.19 & 0.97 \\

   \hline
  & $0.50$   & Naive &   -55.48 & 0.04 & 0.04 & 0.00 & -78.48 & 0.04 & 0.04 & 0.00 \\

   &  & RC\tnote{(2)} \hspace*{0.2cm} (Naive SE) & \phantom{0-}2.97 & 0.10 & 0.09 & 0.87 & \phantom{0-}8.25 & 0.22 & 0.20 & 0.90 \\

   &  & RC\tnote{(2)} \hspace*{0.2cm} (Sandwich) &   \phantom{0}\text{---} & \phantom{0}\text{---} &  0.11 & 0.96 & \phantom{0}\text{---} & \phantom{0}\text{---}  & 0.23 & 0.96 \\

   &  & RC\tnote{(2)} \hspace*{0.2cm} (MI) &    \phantom{0}\text{---} & \phantom{0}\text{---}  & 0.12 & 0.96 & \phantom{0}\text{---} & \phantom{0}\text{---}  & 0.29 & 0.97 \\

     \hline
  & $1.00$   & Naive &       -75.59 & 0.03 & 0.03 & 0.00 & -88.67 & 0.03 & 0.03 & 0.00 \\

   &  & RC\tnote{(2)} \hspace*{0.2cm}  (Naive SE) &    \phantom{0-}3.63 & 0.15 & 0.12 & 0.87 & \phantom{0-}8.76 & 0.32 & 0.28 & 0.90 \\

   &  & RC\tnote{(2)} \hspace*{0.2cm} (Sandwich) &   \phantom{0}\text{---} & \phantom{0}\text{---}  & 0.15 & 0.96 & \phantom{0}\text{---} & \phantom{0}\text{---}  & 0.32 & 0.96 \\ 
  
   &  & RC\tnote{(2)} \hspace*{0.2cm} (MI) &     \phantom{0}\text{---} & \phantom{0}\text{---}  & 0.17 & 0.96 & \phantom{0}\text{---} & \phantom{0}\text{---}  & 0.60 & 0.98 \\

  \hline

\end{tabular}
 \begin{tablenotes}
       \item[(1)]  $\sigma^2=$ the variance of the random, normally distributed covariate measurement error
      \item[(2)] RC = Regression calibration
   \end{tablenotes}
    \end{threeparttable}
\end{table}

\begin{table}[hbp!]
    \centering
    \caption{WHI data analysis (N=$77,805$) results from the Cox Proportional Hazards model for incident diabetes with dietary exposures of energy (kcal/d), protein (g/d), and protein density (\% energy from protein). Results are shown for each stage 2 model fit to the calibrated exposure with naive (model-based) standard errors, sandwich standard errors, and bootstrap standard errors.}
    \label{WHIdataanalysisall}
    \begin{subtable}{\textwidth}
    \begin{adjustbox}{max width=0.9\textwidth}
            \centering
   \begin{threeparttable}[t]
            \caption{Hazard ratio estimates and 95\% confidence intervals (CI) for incident diabetes for a 20\% increase in consumption of energy (kcal/d), protein (g/d), and protein density (\% energy from protein). }\label{WHIdataanalysis1}
   
      \begin{tabular}{llllllllllll}
\hline
\multicolumn{2}{l}{} & \multicolumn{2}{c}{HR (95\% CI)}   \\  \cmidrule(r){3-4} Model\tnote{1} &    Method  & Adjusted for BMI & Not Adjusted for BMI  \\  
\hline
Log-Energy &    RC\tnote{(2)} \hspace*{0.2cm} (Naive SE) &      1.54 (0.94, 2.52) & 2.90 (2.72, 3.08) \\

 &  RC\tnote{(2)} \hspace*{0.2cm} (Sandwich) &  \phantom{--}\text{---}\phantom{-} (0.68, 3.51) & \phantom{--}\text{---}\phantom{-} (2.19, 3.83) \\

 &   RC\tnote{(2)}  \hspace*{0.2cm} (Boot. - Wald)\tnote{(3)} &   \phantom{--}\text{---}\phantom{-} (0, 842640) & \phantom{--}\text{---}\phantom{-} (2.18, 3.86) \\ 
  
 &   RC\tnote{(2)}  \hspace*{0.2cm} (Boot. - Perc)\tnote{(4)} &   \phantom{--}\text{---}\phantom{-} (0.04, 37.17) & \phantom{--}\text{---}\phantom{-} (2.29, 4.03) \\

\hline 
Log-Protein  &    RC\tnote{(2)} \hspace*{0.2cm} (Naive SE) &  1.23 (1.12, 1.34) & 2.12 (2.00, 2.26) \\

 &  RC\tnote{(2)} \hspace*{0.2cm} (Sandwich) &    \phantom{--}\text{---}\phantom{-} (1.08, 1.40) & \phantom{--}\text{---}\phantom{-} (1.60, 2.82) \\

 &   RC\tnote{(2)} \hspace*{0.2cm} (Boot. - Wald)\tnote{(3)}  &  \phantom{--}\text{---}\phantom{-} (1.03, 1.46)  & \phantom{--}\text{---}\phantom{-} (1.59, 2.84) \\ 
&   RC\tnote{(2)} \hspace*{0.2cm} (Boot. - Perc)\tnote{(4)} &  \phantom{--}\text{---}\phantom{-} (1.11, 1.52) & \phantom{--}\text{---}\phantom{-} (1.60, 2.87) \\

 \hline 
Log-Protein &    RC\tnote{(2)} \hspace*{0.2cm} (Naive SE) &    1.64 (1.42, 1.88) & 1.17 (1.02, 1.33) \\ 
 
  Density   &    RC\tnote{(2)} \hspace*{0.2cm} (Sandwich) &    \phantom{--}\text{---}\phantom{-} (1.14, 2.34) & \phantom{--}\text{---}\phantom{-} (0.25, 5.46) \\ 

 &   RC\tnote{(2)} \hspace*{0.2cm} (Boot. - Wald)\tnote{(3)}  &     \phantom{--}\text{---}\phantom{-} (0.83, 3.24) & \phantom{--}\text{---}\phantom{-} (0.19, 6.97) \\ 
 &   RC\tnote{(2)} \hspace*{0.2cm} (Boot. - Perc)\tnote{(4)} &     \phantom{--}\text{---}\phantom{-} (1.31, 3.93) & \phantom{--}\text{---}\phantom{-} (0.28, 8.50)  \\ 

 \hline
\end{tabular}
    \end{threeparttable}
    \end{adjustbox}
    \end{subtable}

    \begin{subtable}{\textwidth}
        \begin{adjustbox}{max width=0.9\textwidth}
            \centering
          \begin{threeparttable}[t]
            \caption{$\beta$ regression parameter estimates and standard errors (SE) estimated by the Cox Proportional Hazards model for incident diabetes for log-energy, log-protein, and log-protein density. }\label{WHIdataanalysis2}
   
      \begin{tabular}{llcccc}
\hline
\multicolumn{2}{l}{} & \multicolumn{2}{c}{Adjusted for BMI} & \multicolumn{2}{c}{Not Adjusted for BMI }   \\  \cmidrule(r){3-4} \cmidrule(r){5-6} Model\tnote{1} &    Method  & $\beta$ & SE &  $\beta$ & SE \\  
\hline
Log-Energy &     RC\tnote{(2)} \hspace*{0.2cm} (Naive SE) &        2.38 & 1.37 & 5.83 & 0.17 \\ 
  
 &  RC\tnote{(2)} \hspace*{0.2cm} (Sandwich) &  \phantom{-}\text{---} & 2.34 & \phantom{-}\text{---} & 0.78 \\ 
 
 &  RC\tnote{(2)}  \hspace*{0.2cm} (Bootstrap) &    \phantom{-}\text{---} & 36.97 & \phantom{-}\text{---} & 0.80  \\ 
\hline 
Log-Protein (g/d) &   RC\tnote{(2)} \hspace*{0.2cm} (Naive SE) &    1.12 & 0.25 & 4.13 & 0.17 \\

 &  RC\tnote{(2)} \hspace*{0.2cm} (Sandwich) &     \phantom{-}\text{---} & 0.36 & \phantom{-}\text{---} & 0.79 \\ 
 
 &   RC\tnote{(2)} \hspace*{0.2cm} (Bootstrap) &   \phantom{-}\text{---} & 0.48 & \phantom{-}\text{---} & 0.82 \\ 
 \hline 
Log-Protein  &   RC\tnote{(2)} \hspace*{0.2cm} (Naive SE) &   2.70 & 0.39 & 0.84 & 0.38 \\ 
 Density &  RC\tnote{(2)} \hspace*{0.2cm} (Sandwich) & \phantom{-}\text{---} & 1.00 & \phantom{-}\text{---} & 4.32 \\ 
 &   RC\tnote{(2)} \hspace*{0.2cm} (Bootstrap) &     \phantom{-}\text{---} & 1.91 & \phantom{-}\text{---} & 5.01 \\
 \hline
\end{tabular}
 \begin{tablenotes}
  \item[(1)] Each model is adjusted for potential confounders and is stratified on age in 5-year categories, hormone therapy trial participation, and DM-C or OS membership
     \item[(2)] RC = Regression calibration
    \item[(3)] Bootstrap with standard normal Wald-based confidence interval
      \item[(4)] Percentile bootstrap confidence interval 
   \end{tablenotes}
    \end{threeparttable}
        \end{adjustbox}
    \end{subtable}%
\end{table}

\begin{table}[hbp!]
    \centering
    \caption{HCHS/SOL data analysis ($N=8,176$) results from the linear regression of baseline systolic blood pressure and the logistic regression of hypertension status each on log-transformed intake of potassium. Results are shown for each stage 2 model fit to the calibrated exposure with naive (model-based) standard errors, sandwich standard errors, and standard errors from the multiple imputation (MI) approach of Baldoni et al. (2021).}
    \label{HCHSdataanalysisall}
    \begin{subtable}{\textwidth}
            \centering
   \begin{threeparttable}[t]
             \caption{Results for a 20\% increase in consumption of potassium are presented. For linear regression, $\log(1.2)\hat{\beta}$ and $95\%$ confidence intervals (CI) are presented. For logistic regression, odds ratio (OR) estimates of $\exp(\log(1.2)\hat{\beta})$ and 95\% CIs are presented. }\label{hchssoldata1}
   
      \begin{tabular}{llcc}
      \hline
\multicolumn{2}{l}{} & Linear Regression & Logistic Regression  \\ Model\tnote{1} &    Method   & $\beta$ ($95\%$ CI) & OR ($95\%$ CI) \\ 
\hline
Log-Potassium  &   RC\tnote{(2)} \hspace*{0.2cm}  (Naive SE) & -0.58 (-1.26, 0.10) & 0.81 (0.63, 1.03) \\

 &  RC\tnote{(2)} \hspace*{0.2cm} (Sandwich) &   \phantom{--}\text{---}\phantom{--} (-1.37, 0.21) & \phantom{--}\text{---}\phantom{--} (0.61, 1.07) \\
 
 &   RC\tnote{(2)} \hspace*{0.2cm} (MI) &    \phantom{--}\text{---}\phantom{--} (-1.78, 0.62) & \phantom{--}\text{---}\phantom{--} (0.52, 1.25) \\

\hline 
\end{tabular}
    \end{threeparttable}
    \end{subtable}

    \begin{subtable}{\textwidth}
            \centering
          \begin{threeparttable}[t]
    \caption{$\beta$ regression parameter estimates and standard errors (SE) estimated by linear and logistic regression models for hypertension-related outcomes for log-potassium.}\label{hchssoldata2}
   
      \begin{tabular}{llcccc}
      \hline
\multicolumn{2}{l}{} & \multicolumn{2}{c}{Linear Regression}  & \multicolumn{2}{c}{Logistic Regression}  \\ Model\tnote{1} &    Method   & $\beta$ & SE  & $\beta$ & SE  \\ 
\hline
Log-Potassium &  RC\tnote{(2)} \hspace*{0.2cm}  (Naive SE) &   -3.17 & 1.90 & -1.17 & 0.68 \\ 
 &  RC\tnote{(2)} \hspace*{0.2cm} (Sandwich) & \phantom{-}\text{---} & 2.21 & \phantom{-}\text{---} & 0.80 \\ 
 &   RC\tnote{(2)} \hspace*{0.2cm} (MI) &     \phantom{-}\text{---} & 3.35 & \phantom{-}\text{---} & 1.22 \\ 

\hline 
\end{tabular}
 \begin{tablenotes}
  \item[(1)] Each model is adjusted for potential confounders 
     \item[(2)] RC = Regression calibration
   \end{tablenotes}
    \end{threeparttable}
    \end{subtable}%
\end{table}

\clearpage

\begin{center}
{\Large \textbf{Supplementary Materials for ``Practical considerations for sandwich variance estimation in two-stage regression settings"} }
\end{center}

\begin{center}
\normalsize{Lillian A. Boe$^{1,*}$, Thomas Lumley$^{2}$, and 
Pamela A. Shaw$^{3}$\\
$^{1}$ Department of Biostatistics, Epidemiology, and Informatics, \\
University of Pennsylvania Perelman School of Medicine, \\
Philadelphia, Pennsylvania, United States
\\
$^{2}$Department of Statistics, Faculty of Science, University of Auckland, \\
Auckland, New Zealand \\
$^{3}$Biostatistics Unit, Kaiser Permanente Washington Health Research Institute \\
Seattle, Washington, United States \\

\textit{*email}: boel@pennmedicine.upenn.edu}

\end{center}

\begin{singlespace}
\section*{Contents}
\end{singlespace}

\setcounter{equation}{0}
\renewcommand{\theequation}{S\arabic{equation}}

\setcounter{section}{0}
\renewcommand{\thesection}{S\arabic{section}}

\setcounter{table}{0}
\renewcommand{\thetable}{S\arabic{table}}

\section{Sufficient Assumptions for Sandwich Variance Estimation}\label{sufficientassumptions}

The stacked estimating equation framework outlined in the prior section may be considered for any regular, asymptotically linear estimators. As illustrated in the main text, this includes any parametric maximum likelihood estimators, as well as the Cox proportional hazards model estimator. This framework does not apply, however, to a range of machine learning models, including Lasso and Random Forest. Here, we outline mild regularity conditions required for the estimators from stage 1 and stage 2 in order to apply the proposed sandwich variance estimator. Specifically, if we assume that the vector of estimated nuisance parameters, $\hat{\boldsymbol{\alpha}}$, from Stage 1 is a regular, asymptotically linear estimator (e.g. a linear regression estimator), the regularity conditions specified by \citet{foutz1977unique} can be used to establish the consistency and uniqueness of this estimator. For the case of a standard maximum likelihood estimator, we may appeal to standard maximum likelihood estimation (MLE) theory to show that $\hat{\boldsymbol{\alpha}}$ is a unique solution to the likelihood equations that is consistent and asymptotically normal \citep{boos2013essential}. Additionally consider an outcome model of interest and suppose all covariates are observed without error. For a generalized linear model, one can appeal to this same standard MLE theory to establish the consistency, uniqueness, and asymptotic normality of the outcome model estimator, $\hat{\boldsymbol{\beta}}$. When the outcome model is a Cox proportional hazards model, the techniques of \citet{andersen1982cox} may be used to establish consistency and asymptotic normality of $\hat{\boldsymbol{\beta}}$. 

We can make sufficient, typical regularity assumptions for our stage 1 model to ensure that we have a consistent and asymptotically normal estimator for $\hat{\boldsymbol{\alpha}}$. A common but not necessary assumption is that $\frac{n}{N} \rightarrow p\in (0,1)$, where $n$ is the number of subjects in the calibration subset. More complex assumptions may also be considered \citep{sarndal2003model}. The regularity of our estimator $\hat{\boldsymbol{\beta}}$ from stage 2 will still hold using a plug-in estimator for $\boldsymbol{\alpha}$ by appealing to Theorem 5.31 in \citet{van2000asymptotic}.

\section{Steps for Computing Bootstrap Standard Errors}\label{bootstrapsteps}

Below we have outlined the steps for computing bootstrap standard errors, as commonly applied in the context of regression calibration, using a stratified bootstrap procedure:

\begin{enumerate}
    \item Choose a number of bootstrap samples to perform (e.g. $B=500$)
    \item For each bootstrap sample,
    \begin{enumerate}
    \item Draw a stratified bootstrap sample with replacement of size $N$: First, draw a sample with replacement of size $n$ from those in the subset only. Then, draw a sample with replacement of size $N-n$ from the non-subset members.
\item Fit the regression calibration (stage 1) model on the bootstrap sample.
\item Use the calibration model fit to the bootstrap sample to get an estimate of the exposure, $\hat{X}_i^{(b)}$, on all main study participants, for $i=1,\ldots,N$.
\item Fit the outcome regression (stage 2) model using the bootstrap sample with $\hat{X}_i^{(b)}$ to obtain an estimate of the $b$th regression model parameters, $\beta^{(b)}$.
    \end{enumerate}
\item Repeat step (2) $B$ times.
\item For an estimate of the adjusted standard error, compute the standard deviation of the $B$ bootstrap estimates of the regression parameter, $\mathbf{\beta}=(\beta^{(1)},...,\beta^{(B)})$.
\end{enumerate} 

\section{Steps for Computing Multiple Imputation-Based Standard Errors}\label{MIsteps}

Below we have outlined the steps for computing standard errors using the resampling-based multiple imputation approach of \citet{baldoni2021use}:

\begin{enumerate}
    \item Choose a number of imputations to perform (e.g. $M=25$)
    \item For each imputation,
    \begin{enumerate}
    \item Draw a bootstrap sample with replacement of size $n$ from those in the subset only. Assuming the calibration subset is a simple random sample of the main study, select individuals in the sample with equal probability.
\item Fit the regression calibration (stage 1) model on the bootstrap sample.
\item Use the calibration model fit to the bootstrap sample to get an estimate of the exposure, $\hat{X}_i^{(m)}$, on all main study participants, for $i=1,\ldots,N$.
\item Fit the outcome regression (stage 2) model using $\hat{X}_i^{(m)}$ and other covariates $Z$ to obtain an estimate of the $m$th regression model parameters, $\beta^{(m)}$.
    \end{enumerate}
\item Repeat step (2) $M$ times.
\item For an estimate of the adjusted standard error of $\hat{\beta}$, compute $\hat{V}^*=\frac{1}{M}\sum_{m=1}^M \hat{V}^{(m)}+\frac{1}{M-1}\sum_{m=1}^M\Big(\hat{\beta}^{(m)}-\bar{\hat{\beta}}\Big)^2$, where $\bar{\hat{\beta}}=\frac{1}{M}\sum_{m=1}^M\hat{\beta}^{(m)}$ and $\hat{\beta}^{(m)}$ and $\hat{V}^{(m)}$ represent the estimated regression coefficient and its estimated variance, respectively, using the $m$-th completed data set.
\end{enumerate}

\citet{baldoni2021use} also considered robust estimators for the mean and standard deviation used to compute the adjusted variance in step 4. Specifically, for robustness to skewed estimates, the median and median absolute deviation were used. Further details on the multiple imputation-based variance estimator are described in \citet{baldoni2021use} and code for implementing this procedure is available on GitHub at \url{https://github.com/plbaldoni/HCHSsim}. \\~\

\section{Details on Simulation Study to Illustrate Performance of Sandwich Variance Estimator}\label{simulationstudydetials}

We conducted a simulation study to show how the sandwich variance approach and competing estimators perform under various different settings. We simulate two covariates, $X_i$ and $Z_i$, from a multivariate normal (MVN) distribution with mean 0 (i.e. $\mu=[0, 0]$) and a covariance matrix with all diagonal elements equal to 1. We vary the off-diagonal elements between 0.3 and 0.7 to represent low and high correlation, respectively, between the two covariates, i.e.

\begin{equation*}
    \Sigma=\begin{bmatrix}
1 & r \\
r & 1 \\ 
\end{bmatrix},
\end{equation*}

\noindent where $r\in(0.3, 0.7)$. Next, we assume the logistic regression model, $p_i=P(Y_i=1)=\frac{\exp(\beta_0+\beta_X X_i + \beta_Z Z_i)}{1+\exp(\beta_0+\beta_X X_i + \beta_Z Z_i)}$ and fix  $\beta_0=0.2$, $\beta_X=\log(1.5)$, and $\beta_Z=\log(0.7)$. In epidemiologic settings, $\beta_X=\log(1.5)$ represents a log-odds ratio corresponding to the exposure of interest of a moderate size. To simulate the binary outcome $Y_i$, we generated $N$ variables $\textrm{Un}_i$ from a Uniform(0,1) distribution and then let $Y_i=1$ if $\textrm{Un}_i<p_i$ and 0 otherwise. 

Later, we run a set of simulations for the random sample case designed to assess the performance of the proposed sandwich variance estimator when a Cox proportional hazards model is considered as our stage 2 model. To simulate this outcome, we generated event times from a continuous time exponential distribution with parameter $\lambda=0.23 \exp(\beta_X X_i + \beta_Z Z_i)$, where $\beta_X=\log(1.5)$ and $\beta_Z=\log(0.7)$. We let our censoring time be 2 such that any subject who experienced the event prior to this time was assumed to have experienced the event of interest. Assigning these parameters resulted in an observed event rate that approximated that for our logistic regression outcome model, roughly 38\%.

Our error-prone covariate $X_i^*$ is simulated to represent a hypothetical dietary intake variable using the following linear measurement error model, $X_i^*=\delta_0 + \delta_1 X_i+\delta_2 Z_i +e_i$, where $\delta_X=0.20$, $\delta_X=0.37$ and $\delta_Z=0.15$. We let $e_i \sim N(0,\sigma^2)$  and considered $\sigma^2$ values of 0, 0.25, 0.50, and 1.00 to represent cases of zero, low, moderate, and high measurement error. We assume that our hypothetical biomarker subset is a random sample of $n=450$ subjects from the main study. Our simulated hypothetical biomarker of interest, $X^{**}_i$, is generated to follow the classical measurement error model $X^{**}_i=X_i+\epsilon_i$, and we let $\epsilon \sim N(0,0.2)$.

For our simulation studies designed to mimic the structure the complex survey design data setting, we used the simulation pipeline proposed by \citet{baldoni2021use}, which simulates a superpopulation using a three stage design and then draws samples from it using a stratified complex survey sampling scheme. The resulting simulated data sets included sampling weights, stratification variables, and cluster indicators. To simulate $X_i$ and $Z_i$ such that each stratum and cluster had a slightly different covariate distribution, we began by considering four different mean vectors and covariance matrices for a multivariate normal distribution, i.e. $\mu_1=(0.15, 0.15)$, $\mu_2=(-0.15, -0.15)$, $\mu_3=(0.3, 0.3)$, $\mu_3=(-0.3, -0.3)$, $\Sigma_1=\Sigma+0.15*\Sigma$, $\Sigma_2=\Sigma-0.15*\Sigma$, $\Sigma_3=\Sigma+0.3*\Sigma$, and $\Sigma_4=\Sigma-0.3*\Sigma$, where $\Sigma$ is defined above for simulations from a simple random sample. Within each stratum $s$, we created a small covariate difference for individuals in a certain block group $g$ by simulating variables $\omega_{g1}$, $\omega_{g2}$, $\omega_{g3}$ and $\omega_{g4}$ from a Uniform($-0.015,0.015$), Uniform($-0.0225,0.0225$), Uniform($-0.03,0.03$), and Uniform($-0.045, 0.045$) distribution, respectively. Similarly, we generated variables $\rho_{gs}$ from a Uniform($-0.15*{r}_s,0.15*{r}_s$)  distribution for $s=1\ldots,4$, where ${r}_s$ is the off-diagonal element of $\Sigma_s$. Then, the covariate for an individual in block group $g$ and stratum $s$ was simulated from a MVN$(\mu_s+\omega_{gs},\Sigma_s+\rho_{gs})$ distribution. All other data generation settings besides the sampling structure and covariate simulation were kept the same as in the simple random sampling case, including the simulation of $X_i^{*}$, $X_i^{**}$, and $Y_i$. In a complex survey sampling scheme of this type, it was not possible to fix the total number of individuals selected for a simulated sample exactly, but we were able to obtain sample sizes of approximately $N=1000$ and $N=10,000$. 

For all settings studied, we conducted $1000$ simulation iterations. Simulations for the simple random sample which computed standard error estimates using the bootstrap used $B=500$ bootstrap samples. Confidence intervals in Tables 1 and S1 are constructed using the typically applied Wald confidence interval computed using bootstrap standard errors. In Table S2, we used 500 bootstrap samples for the Wald and percentile confidence intervals to mimic Table 1, but 1000 bootstrap samples for the bias-corrected and accelerated (BCA) bootstrap interval, as required for computation in a data set with $N=1000$.  For simulations from the complex survey design, we used $M=25$ imputations when applying the multiple imputation based procedure of \cite{baldoni2021use}. 

\section{Supplementary Details on the WHI Data Analysis}\label{WHIsupplement}

To fit the stage 1 model in the WHI data analysis, we modified calibration models that were previously developed for self-reported intakes of energy, protein, and protein density  by \citet{neuhouser2008use} and used by \citet{tinker2011biomarker} to obtain incident diabetes hazard ratios in the WHI cohort. Specifically, one recommended approach for regression calibration approach to avoid potential bias is to include in the calibration model the same set of covariates as the outcome model \citep{kipnis2009modeling}. Thus, we expanded the set of covariates from the former calibration models to include all confounding variables that will be included in our outcome model. All stage 1 models therefore included body mass index (BMI), age, race-ethnicity, income, education, physical activity in units of metabolic equivalent tasks per week, smoking status, alcohol consumption, hypertension, history of cardiovascular disease, family history of diabetes, and hormone use. We note of particular specific concern is that bias will arise in the outcome model coefficients if there are covariates associated with outcome that are in the calibration model that are left out of the outcome model; or similarly, if there were important covariates left out of the calibration model that were included in the outcome model.
See further discussion in \citep{kipnis2009modeling} regarding conditions that allow extra covariates in the calibration model, known as precision covariates, such as when those covariates have no association with the outcome. 

Incident diabetes in the WHI was recorded using a self-reported questionnaire at annual follow-up visits. As in \cite{tinker2011biomarker}, we consider the Cox proportional hazards model for our stage 2 model, stratified on age in 5-year categories, hormone therapy trial participation, and DM-C or OS membership. All stage 2 models for diabetes were adjusted by the same set of confounders included in the stage 1 models. We also stratified our stage 2 models on age in 5-year categories, hormone therapy trial participation, and DM-C or OS membership. Following \citet{tinker2011biomarker}, we fit two versions of the stage 2 model, one which is adjusted for BMI and the other which is not. This issue, which is discussed by \cite{tinker2011biomarker}, relates to the fact that BMI may be a mediator of the relationship between energy intake and diabetes. We compare model-based standard errors (naive SE) to those estimated by the proposed sandwich estimator introduced in the main paper and the standard bootstrap procedure using $B=500$ bootstrap samples.

We consider data from women who participated in either the comparison arm of the Dietary Modification trial (DM-C) or the Observational Study (OS) in the WHI \citep{ritenbaugh2003women,langer2003women}. Note that neither women from the DM-C nor the OS received study interventions. By adopting the same exclusion criteria described by \citet{tinker2011biomarker}, we obtained a final analytic data set of $77,805$ participants. These criteria essentially attempt to align the characteristics of participants in the DM-C and OS cohorts as well as exclude any women with missing data or who reported diabetes at baseline. In our analysis, baseline was defined as the time of the first self-reported dietary assessment post-enrollment, year 1 for the DM-C and year 3 for the OS. 

Following \citet{tinker2011biomarker}, we excluded women who reported diabetes at baseline or during the first year of follow-up for the comparison arm of the WHI Dietary Modification trial (DM-C) participants ($n=724$) or the first three years of follow-up for the WHI Observational Study (OS) participants ($n=4109$). In an attempt to align characteristics of women in the DM-C trial with those of women in the OS, the following women in OS were also excluded: those who had breast, colorectal, or other cancer within 10 years prior to enrollment ($n=8677)$, stroke or myocardial infarction within 6 months prior to enrollment ($n=155$), body mass index (BMI) $<18$ ($n=678$), hypertension (systolic blood pressure $>200$ or diastolic blood pressure $>105$)($n=244$), reported daily energy intake of $<600$ kcal or $>5000$ kcal ($n=3571$), $\geq 10$ meals prepared away from home each week ($n=3598$), a special low-fiber diet ($n=568$), a special malabsorption-related diet ($n=514$), inadvertent weight loss of $>15$ pounds within 6 months of enrollment ($n=594$), and diabetes diagnosis recorded before age 21 at enrollment ($n=95$). After applying these criteria and including only the participants with no missing data from the stage 1 and stage 2 model variables, we obtained our analytic cohort with $77,805$ participants. Of these $77,805$ women, $19,945$ (25.6\%) were from the DM-C and $57,860$ (74.4\%) were from the OS. Our stage 1 models included 356 eligible women from the Nutritional Biomarker Study who did not have missing data.

Our estimated hazard ratios for all models were somewhat different than those reported by \citet{tinker2011biomarker}. Specifically, when BMI was excluded from the outcome model, \citet{tinker2011biomarker} showed that a HR (95\% CI) of 2.41 (2.06, 2.82) was associated with a 20\% increase in energy intake, which differs slightly from our 2.90 (2.18, 3.86) with adjusted standard errors estimated by the bootstrap. For the BMI-adjusted analyses, \citet{tinker2011biomarker} reported that a HR (95\% CI) of 1.30 (0.96, 1.76) for a 20\% increase in energy, compared to our 1.54 (0, 842640) with bootstrap standard errors. Note that, in constrast to our results for this model,  \citet{tinker2011biomarker} did not obtain an unstable bootstrap standard error estimate. These differences may be explained by a few discrepancies in our calibration model and our analytic cohort, including the final analysis sample size for the stage 1 and stage 2 models. Specifically, even after applying the same exclusion criteria and considering only participants with complete data, we had a slightly different data set with $19,968$ DM-C and $57,860$ OS participants, compared to \citet{tinker2011biomarker} whose analytic data set contained $19,111$ DM-C and $55,044$ OS participants for a total of $74,155$ in the analysis. One major difference between our analysis approach and that of \citet{tinker2011biomarker} related to the variables included in the stage 1 and stage 2 models. As noted above, we chose to include the same set of confounders, $Z$, in the stage 1 and stage 2 models, while \citet{tinker2011biomarker} included a set of confounders in the stage 2 outcome model that were not included in the stage 1 model. Due to the expanded stage 1 model, we had a smaller calibration data set with complete covariates, which could also have contributed to the observed instability of the bootstrap in our example. Finally, we note that \citet{tinker2011biomarker} included the variables glycemic index and glycemic load in their stage 2 models. We chose to exclude these variables from all models due to numerical instability that they created when added to the stage 1 models, potentially due to correlation with other variables. 

\section{Supplementary Details on the HCHS/SOL Data Analysis}\label{hchssupplement}

In our reanalysis of the HCHS/SOL data, we fit our stage 1 model to $n=310$ SOLNAS participants, excluding any SOLNAS participants who were ineligible for the stage 2 analysis from having had a previous diagnosis of high blood pressure or hypertension or making use of antihypertensive medication. This sample also excluded 11 SOLNAS participants who had unreliable (biomarker or self-reported values that were too extreme) for sodium.  Our stage 2 analysis of the HCHS/SOL data included $8,176$ participants from the original HCHS/SOL cohort ($N=16,415$). We started with the subset used by \citet{baldoni2021use}, which was constructed by taking a random sample of $8,208$ HCHS/SOL participants and excluding $83$ participants who had missing covariate data. For our sample, we considered these $8,176$ participants with complete data, then added back the remaining 51 eligible SOLNAS ancillary study participants who were not selected by the original random sample.

While the convention in the HCHS/SOL study has previously been to ignore the survey design in the fitting of the stage 1 model, we chose to account for the design in the model, a decision that was made to capture the variability in the complex survey design. To account for the survey design in the stage 1 model, we subset the parent cohort HCHS/SOL survey design to the participants in the SOLNAS substudy. As described in the main manuscript, attention must be paid to assigning the strata when using data from a complex survey design. For this analysis, we specified the strata as the cross-classification of the subset indicator, $V_i$, with the strata variables from the HCHS/SOL design. Since SOLNAS is not a nested subset by design, special consideration was required to determine the strata to be used in this data example. The parent HCHS/SOL study included multi-level strata, where the top level was field center, while SOLNAS was only stratified by field center. Thus, our best approximation for determining the strata in this data example was to use the full set of multi-level strata for those not in the SOLNAS subset ($V_i=0$) and just the field center for those in the SOLNAS subset ($V_i=1$).

The confounding variables of interest, $Z_i$ used in our stage 1 and stage 2 models are body mass index (BMI), age, Hispanic/Latino background, income, education, physical activity, smoking status, alcohol consumption, field center, language preference, sex, nativity, family history of cardiovascular health disease, and hypercholesterolemia. All analyses used log-transformed biomarkers for sodium and potassium and  log-transformed self-reported 24-hour recall measures. The outcome variables of hypertension and systolic blood pressure were recorded at the HCHS/SOL baseline, in-person clinical examination visit (2008-2011). We consider standard errors estimated by the model (Naive SE), the sandwich, and the MI procedure with $M=25$ imputations.

\section{Sandwich Variance Example: Regression Calibration Applied to Logistic Regression}\label{sandwichexample}

In this section, we use an example to illustrate how one might derive the stacked estimating equations, $U_i(\theta)$, and $A(\hat{\theta})$ and $B(\hat{\theta})$ matrices in order to obtain a sandwich variance estimator. For this example, we consider the setting in which regression calibration is applied to a logistic regression outcome model. Suppose we are interested in a binary outcome, such as hypertension. Consider a logistic regression outcome model such that for subjects $i=1,...,N$, we observe $\mathbf{Y}=(Y_1,...,Y_N)$. Define  $Y_i$ as 1 if the $i^{th}$ individual has hypertension and 0 otherwise. Denote our unknown regression parameter vector as $\boldsymbol{\beta}=[\beta_0,\beta_X,\beta_Z]$. 
Suppose we are interested in fitting the following stage 2 model relating the probability of hypertension with our estimated exposure $\hat{X}_i$ and other covariates of interest $Z_i$:

\begin{equation}\label{logisticregcalibrated}
    \mathrm{logit}\, [P(Y=1|\hat{X}(\hat{\boldsymbol{\alpha}}),Z;\boldsymbol{\beta})]={\beta}_0+{\beta}_X^T \hat{X}_{i}+{\beta}_Z^T  Z_{i}.
\end{equation}

For the case of a logistic regression outcome model, our parameter vector of interest, which includes the nuisance parameters from the stage 1 model and our stage 2 outcome regression model parameters is then: $\theta=(\alpha_0,\alpha_X,\alpha_Z,\beta_0,\beta_X,\beta_Z)$. We can then use the M-estimation approach of \citet{boos2013essential} to obtain the vector of stacked estimating equations for the parameter vector $\theta$ as follows:

\begin{equation}
    U_i(\theta)= \begin{bmatrix}[1.5]  U_{i1}(\theta) \\   U_{i2}(\theta)  \end{bmatrix}= 
    \begin{bmatrix}[1.5] U_{i1(a)}(\theta) \\   U_{i1(b)}(\theta) \\ U_{i1(c)}(\theta) \\ U_{i2(a)}(\theta) \\ U_{i2(b)}(\theta) \\ U_{i2(c)}(\theta)  \end{bmatrix}
= \begin{bmatrix}[1.5]
    V_i(X_i^{**}-\alpha_0-\alpha_X X_i^{*}-\alpha_Z Z_i) \\ 
    V_i(X_i^{**}-\alpha_0-\alpha_X X_i^{*}-\alpha_Z Z_i)X_i^{*} \\ 
    V_i(X_i^{**}-\alpha_0-\alpha_X X_i^{*}-\alpha_Z Z_i)Z_i\\ 
    Y_i-p_i \\ 
    \hat{X}_{i}(Y_i-p_i) \\ 
    Z_{i}(Y_i-p_i) \\ 
    \end{bmatrix}
\end{equation}

\noindent Note that since $\hat{X}_i=E(X_i^{**} | X_i^*,Z_i)=\alpha_0+\alpha_X X_i^* +\alpha_Z Z_i$, we have:

\begin{equation*}
    U_i(\theta)=
    \begin{bmatrix}[1.5]
 V_i(X_i^{**}-\alpha_0-\alpha_X X_i^{*}-\alpha_Z Z_i) \\ 
    V_i(X_i^{**}-\alpha_0-\alpha_X X_i^{*}-\alpha_Z Z_i)X_i^{*} \\ 
    V_i(X_i^{**}-\alpha_0-\alpha_X X_i^{*}-\alpha_Z Z_i)Z_i\\ 
    Y_i-p_i \\ 
    (\alpha_0+\alpha_X X_i^* +\alpha_Z Z_i)(Y_i-p_i) \\ 
    Z_{i}(Y_i-p_i) \\ 
    \end{bmatrix}
\end{equation*} 

\noindent As described in the main manuscript, the estimates $\hat{\theta}$ can be found by solving the equations $\sum_{i=1}^NU_i(\theta)=0$. A sandwich estimator for the variance of $\hat{\theta}$ can then be obtained as:

\begin{equation}
    V(\hat{\theta})=A(\hat{\theta})^{-1} B(\hat{\theta}) \left[A(\hat{\theta})^{-1}\right]^T
\end{equation}

\noindent where

\begin{equation}
    A(\hat{\theta})=\sum_{i=1}^N \frac{\partial U_i(\theta)}{\partial \theta} |_{\theta=\hat{\theta}}
\end{equation}

\noindent and 

\begin{equation}
    B(\hat{\theta})=\sum_{i=1}^N U_i(\theta)U_i(\theta)^T
\end{equation}

\noindent For our example where regression calibration is applied using a linear (stage 1) model to a logistic regression (stage 2) outcome model, we can explicitly define these matrices as follows: 

\begin{landscape}

\begin{equation*}
A(\theta)=\sum_{i=1}^N\begin{bmatrix}
\frac{\partial U_{i1(a)}(\theta)}{\partial \alpha_0 } & 
\frac{\partial U_{i1(a)}(\theta)}{\partial \alpha_{X}} & 
\frac{\partial U_{i1(a)}(\theta)}{\partial  \alpha_{Z}} & 
\frac{\partial U_{i1(a)}(\theta)}{\partial  \beta_{0}} &
\frac{\partial U_{i1(a)}(\theta)}{\partial  \beta_{X}} &
\frac{\partial U_{i1(a)}(\theta)}{\partial  \beta_{Z}} \\
\frac{\partial U_{i1(b)}(\theta)}{\partial  \alpha_{0} } & 
\frac{\partial U_{i1(b)}(\theta)}{\partial \alpha_X} & 
\frac{\partial U_{i1(b)}(\theta)}{\partial  \alpha_{Z}} & 
\frac{\partial U_{i1(b)}(\theta)}{\partial  \beta_{0}} &
\frac{\partial U_{i1(b)}(\theta)}{\partial \beta_{X}} &
\frac{\partial U_{i1(b)}(\theta)}{\partial  \beta_{Z}} \\
\frac{\partial U_{i1(c)}(\theta)}{\partial  \alpha_{0} } & 
\frac{\partial U_{i1(c)}(\theta)}{\partial  \alpha_X} & 
\frac{\partial U_{i1(c)}(\theta)}{\partial  \alpha_{Z}} & 
\frac{\partial U_{i1(c)}(\theta)}{\partial \beta_{0}} &
\frac{\partial U_{i1(c)}(\theta)}{\partial \beta_{X}} &
\frac{\partial U_{i1(c)}(\theta)}{\partial  \beta_{Z}} \\
\frac{\partial U_{i2(a)}(\theta)}{\partial \alpha_{0} } & 
\frac{\partial U_{i2(a)}(\theta)}{\partial \alpha_X} & 
\frac{\partial U_{i2(a)}(\theta)}{\partial \alpha_{Z}} & 
\frac{\partial U_{i2(a)}(\theta)}{\partial \beta_0} &
\frac{\partial U_{i2(a)}(\theta)}{\partial  \beta_{X}} &
\frac{\partial U_{i2(a)}(\theta)}{\partial  \beta_{Z}} \\
\frac{\partial U_{i2(b)}(\theta)}{\partial \alpha_{0} } & 
\frac{\partial U_{i2(b)}(\theta)}{\partial  \alpha_X} & 
\frac{\partial U_{i2(b)}(\theta)}{\partial  \alpha_{Z}} & 
\frac{\partial U_{i2(b)}(\theta)}{\partial \beta_0} &
\frac{\partial U_{i2(b)}(\theta)}{\partial \beta_{X}} &
\frac{\partial U_{i2(b)}(\theta)}{\partial  \beta_{Z}} \\
\frac{\partial U_{i2(c)}(\theta)}{\partial \alpha_{0} } & 
\frac{\partial U_{i2(c)}(\theta)}{\partial  \alpha_X} & 
\frac{\partial U_{i2(c)}(\theta)}{\partial  \alpha_{Z}} & 
\frac{\partial U_{i2(c)}(\theta)}{\partial  \beta_0} &
\frac{\partial U_{i2(c)}(\theta)}{\partial \beta_{X}} &
\frac{\partial U_{i2(c)}(\theta)}{\partial  \beta_{Z}} 

\end{bmatrix}
\end{equation*}

\scriptsize
\begin{equation*}
=\sum_{i=1}^N\begin{bmatrix}
-V_i & -X_i^* V_i & -Z_i V_i & 0 & 0 & 0 \\ 
-X_i^* V_i & -{X_i^*}^2V_i & -X_i^* Z_i V_i & 0 & 0 & 0 \\
-Z_i V_i & -X_i^* Z_i V_i & -Z_i^2 V_i & 0 & 0 & 0 \\ 
-\beta_X p_i(1-p_i) & -\beta_X X_i^* p_i(1-p_i) & -\beta_X Z_i p_i(1-p_i) & -p_i(1-p_i) & -\hat{X}_i p_i(1-p_i) & -Z_i p_i(1-p_i) \\ \hat{X}_i(\beta_X p_i^2-\beta_X p_i)+Y_i-p_i  & \hat{X}_i(\beta_X  X_i^*p_i^2-\beta_X  X_i^* p_i)+X_i^*(Y_i-p_i)
& \hat{X}_i(\beta_X  Z_i p_i^2-\beta_X  Z_i  p_i)+Z_i (Y_i-p_i) & - \hat{X}_i p_i(1-p_i) & - (\hat{X}_i)^2 p_i(1-p_i) & - \hat{X}_i Z_i  p_i(1-p_i) \\ -\beta_X Z_i p_i(1-p_i) & -\beta_X Z_i X_i^* p_i(1-p_i) & -\beta_X Z_i^2 p_i(1-p_i) & - Z_i p_i(1-p_i) & - Z_i(\hat{X}_i) p_i(1-p_i) & - Z_i^2 p_i(1-p_i)
\end{bmatrix}
\end{equation*}

\normalsize

and

\begin{equation*}
B(\theta) =\sum_{i=1}^N \begin{bmatrix} (U_{i1(a)}(\theta))^2 & U_{i1(a)}(\theta) U_{i1(b)}(\theta) & U_{i1(a)}(\theta) U_{i1(c)}(\theta) & U_{i1(a)}(\theta) U_{i2(a)}(\theta) &
U_{i1(a)}(\theta) U_{i2(b)}(\theta) & U_{i1(a)}(\theta) U_{i2(c)}(\theta) \\
U_{i1(b)}(\theta) U_{i1(a)}(\theta) & (U_{i1(b)}(\theta))^2 & U_{i1(b)}(\theta)U_{i1(c)}(\theta) & U_{i1(b)}(\theta)U_{i2(a)}(\theta) &
U_{i1(b)}(\theta)U_{i2(b)}(\theta) & U_{i1(b)}(\theta)U_{i2(c)}(\theta) \\
U_{i1(c)}(\theta)U_{i1(a)}(\theta) & U_{i1(c)}(\theta)U_{i1(b)}(\theta) & (U_{i1(c)}(\theta))^2 & U_{i1(c)}(\theta)U_{i2(a)}(\theta) &
U_{i1(c)}(\theta)U_{i2(b)}(\theta) & U_{i1(c)}(\theta)U_{i2(c)}(\theta) \\
U_{i2(a)}(\theta)U_{i1(a)}(\theta) & U_{i2(a)}(\theta)U_{i1(b)}(\theta) & U_{i2(a)}(\theta)U_{i1(c)}(\theta) &  (U_{i2(a)}(\theta))^2 &
U_{i2(a)}(\theta)U_{i2(b)}(\theta) & U_{i2(b)}(\theta)U_{i2(c)}(\theta) \\
U_{i2(b)}(\theta)U_{i1(a)}(\theta) & U_{i2(b)}(\theta)U_{i1(b)}(\theta) & U_{i2(b)}(\theta)U_{i1(c)}(\theta) &  U_{i2(b)}(\theta)U_{i2(a)}(\theta) &
(U_{i2(b)}(\theta))^2 & U_{i2(c)}(\theta)U_{i2(c)}(\theta) \\
U_{i2(c)}(\theta)U_{i1(a)}(\theta) & U_{i2(c)}(\theta)U_{i1(b)}(\theta) & U_{i2(c)}(\theta)U_{i1(c)}(\theta) &  U_{i2(c)}(\theta)U_{i2(a)}(\theta) &
 U_{i2(c)}(\theta)U_{i2(b)}(\theta) &
 (U_{i2(c)}(\theta))^2 

\end{bmatrix} 
\end{equation*}

\begin{equation*}
   =\sum_{i=1}^N \begin{bmatrix} 
    V_i(X_i^{**}-\alpha_0-\alpha_X X_i^{*}-\alpha_Z Z_i) \\ 
    V_i(X_i^{**}-\alpha_0-\alpha_X X_i^{*}-\alpha_Z Z_i)X_i^{*} \\ 
    V_i(X_i^{**}-\alpha_0-\alpha_X X_i^{*}-\alpha_Z Z_i)Z_i\\ 
    Y_i-p_i \\ 
    \hat{X}_{i}(Y_i-p_i) \\ 
    Z_{i}(Y_i-p_i) \\ 
    \end{bmatrix}
    \begin{bmatrix} 
     V_i(X_i^{**}-\alpha_0-\alpha_X X_i^{*}-\alpha_Z Z_i) \\ 
    V_i(X_i^{**}-\alpha_0-\alpha_X X_i^{*}-\alpha_Z Z_i)X_i^{*} \\ 
    V_i(X_i^{**}-\alpha_0-\alpha_X X_i^{*}-\alpha_Z Z_i)Z_i\\ 
    Y_i-p_i \\ 
    \hat{X}_{i}(Y_i-p_i) \\ 
    Z_{i}(Y_i-p_i) \\ 
    \end{bmatrix} ^T
\end{equation*}

\end{landscape}

\begin{table}
\centering
      \begin{threeparttable}[t]
            \caption{The median percent (\%) bias, median standard errors (ASE), empirical median absolute deviation (MAD) and coverage probabilities (CP) for 1000 simulated data sets from a simple random sample for a Cox proportional hazards stage 2 model (observed event rate = 0.38)  fit to true exposure, naive exposure, and calibrated exposure with naive (model-based) standard errors,  sandwich standard errors, and  bootstrap standard errors. We let $\beta=\log(1.5)$ and vary the correlation between  $X$ and $Z$, the sample size ($N$), and the measurement error variance ($\sigma^2$). Sample size of the calibration subset is $n=450$.}\label{table1appendixsims}
   
        \begin{tabular}{lllcccccccc}

\hline
\multicolumn{3}{l}{} & \multicolumn{4}{c}{Low Correlation}  & \multicolumn{4}{c}{High Correlation}  \\ \cmidrule(r){4-7} \cmidrule{8-11}  $N$ & $\sigma^2$\tnote{(1)} & Method   & \% Bias & MAD & ASE & CP & \% Bias & MAD & ASE & \multicolumn{1}{c}{CP}   \\ 

\hline
 $1000$ & $0.00$ & Truth & \phantom{0-}0.10 & 0.05 & 0.05 & 0.94 & \phantom{0-}0.21 & 0.07 & 0.07 & 0.94 \\

\hline
  & $0.25$  & Naive &  
  -12.36 & 0.08 & 0.08 & 0.91 & -42.22 & 0.09 & 0.09 & 0.52 \\

   &  & RC\tnote{(2)} \hspace*{0.2cm} (Naive SE) &    \phantom{0}-2.06 & 0.09 & 0.09 & 0.94 & \phantom{0}-2.54 & 0.15 & 0.15 & 0.95 \\

   &  & RC\tnote{(2)} \hspace*{0.2cm} (Sandwich) &    \phantom{0}\text{---} &\phantom{0}\text{---} & 0.09 & 0.95 & \phantom{0}\text{---} &\phantom{0}\text{---} & 0.16 & 0.95 \\

   &  & RC\tnote{(2)} \hspace*{0.2cm} (Bootstrap) &      \phantom{0}\text{---} &\phantom{0}\text{---} & 0.10 & 0.95 & \phantom{0}\text{---} &\phantom{0}\text{---} & 0.16 & 0.96 \\

   \hline
  
  & $0.50$  & Naive & -48.12 & 0.07 & 0.06 & 0.16 & -67.79 & 0.07 & 0.07 & 0.02 \\ 

   &  & RC\tnote{(2)} \hspace*{0.2cm}  (Naive SE) &     \phantom{0}-2.88 & 0.12 & 0.12 & 0.94 & \phantom{0}-3.04 & 0.21 & 0.20 & 0.95 \\

   &  & RC\tnote{(2)} \hspace*{0.2cm} (Sandwich) &     \phantom{0}\text{---} &\phantom{0}\text{---} & 0.12 & 0.95 & \phantom{0}\text{---} &\phantom{0}\text{---} & 0.21 & 0.96 \\

   &  & RC\tnote{(2)} \hspace*{0.2cm} (Bootstrap) &   \phantom{0}\text{---} &\phantom{0}\text{---} & 0.13 & 0.95 & \phantom{0}\text{---} &\phantom{0}\text{---} & 0.22 & 0.96 \\

  \hline
  & $1.00$  & Naive &       -71.15 & 0.05 & 0.05 & 0.00 & -82.64 & 0.05 & 0.05 & 0.00 \\ 
 
   &  & RC\tnote{(2)} \hspace*{0.2cm}  (Naive SE) &   \phantom{0}-2.90 & 0.16 & 0.16 & 0.94 & \phantom{0}-4.26 & 0.29 & 0.28 & 0.95 \\

   &  & RC\tnote{(2)} \hspace*{0.2cm} (Sandwich) &   \phantom{0}\text{---} &\phantom{0}\text{---} & 0.17 & 0.95 & \phantom{0}\text{---} &\phantom{0}\text{---} & 0.29 & 0.97 \\

   &  & RC\tnote{(2)} \hspace*{0.2cm} (Bootstrap) &    \phantom{0}\text{---} &\phantom{0}\text{---} & 0.17 & 0.96 & \phantom{0}\text{---} &\phantom{0}\text{---} & 0.32 & 0.98 \\

      \hline
 $10000$ & $0.00$ & Truth &   \phantom{0}-0.03 & 0.02 & 0.02 & 0.96 & \phantom{0}-0.29 & 0.02 & 0.02 & 0.96 \\ 
 
 \hline

  &  $0.25$ & Naive &    -12.42 & 0.03 & 0.03 & 0.52 & -42.27 & 0.03 & 0.03 & 0.00 \\ 
  
   &  & RC\tnote{(2)}  \hspace*{0.2cm}   (Naive SE) &    \phantom{0}-2.81 & 0.04 & 0.03 & 0.82 & \phantom{0}-2.38 & 0.06 & 0.05 & 0.87 \\

   &  & RC\tnote{(2)} \hspace*{0.2cm} (Sandwich) &   \phantom{0}\text{---} &\phantom{0}\text{---} & 0.04 & 0.94 & \phantom{0}\text{---} &\phantom{0}\text{---} & 0.06 & 0.96 \\

   &  & RC\tnote{(2)} \hspace*{0.2cm} (Bootstrap) &   \phantom{0}\text{---} &\phantom{0}\text{---} & 0.04 & 0.95 & \phantom{0}\text{---} &\phantom{0}\text{---} & 0.07 & 0.96 \\

   \hline

  & $0.50$  & Naive &   -47.64 & 0.02 & 0.02 & 0.00 & -67.70 & 0.02 & 0.02 & 0.00 \\

   &  & RC\tnote{(2)} \hspace*{0.2cm} (Naive SE) &    \phantom{0}-3.32 & 0.05 & 0.04 & 0.81 & \phantom{0}-2.88 & 0.08 & 0.06 & 0.85 \\

   &  & RC\tnote{(2)} \hspace*{0.2cm} (Sandwich) & \phantom{0}\text{---} &\phantom{0}\text{---} & 0.05 & 0.95 & \phantom{0}\text{---} &\phantom{0}\text{---} & 0.09 & 0.96 \\ 
 
   &  & RC\tnote{(2)} \hspace*{0.2cm} (Bootstrap) &     \phantom{0}\text{---} &\phantom{0}\text{---} & 0.06 & 0.95 & \phantom{0}\text{---} &\phantom{0}\text{---} & 0.09 & 0.96 \\

     \hline

  & $1.00$  & Naive &     -70.94 & 0.01 & 0.02 & 0.00 & -82.85 & 0.01 & 0.02 & 0.00 \\ 
  
   &  & RC\tnote{(2)} \hspace*{0.2cm}  (Naive SE) &    \phantom{0}-3.22 & 0.08 & 0.05 & 0.81 & \phantom{0}-2.77 & 0.12 & 0.09 & 0.86 \\

   &  & RC\tnote{(2)} \hspace*{0.2cm} (Sandwich) &  \phantom{0}\text{---} &\phantom{0}\text{---} & 0.07 & 0.95 & \phantom{0}\text{---} &\phantom{0}\text{---} & 0.12 & 0.96 \\

   &  & RC\tnote{(2)} \hspace*{0.2cm} (Bootstrap) &      \phantom{0}\text{---} &\phantom{0}\text{---} & 0.08 & 0.95 & \phantom{0}\text{---} &\phantom{0}\text{---} & 0.14 & 0.97 \\

  \hline

\end{tabular}
 \begin{tablenotes}
     \item[(1)]  $\sigma^2=$ the variance of the random, normally distributed covariate measurement error
      \item[(2)] RC = Regression calibration
   \end{tablenotes}
    \end{threeparttable}
\end{table}

\begin{table}
\centering
      \begin{threeparttable}[t]
            \caption{The median percent (\%) bias, median standard errors (ASE), empirical median absolute deviation (MAD) and coverage probabilities (CP) for 1000 simulated data sets from a simple random sample for a logistic regression stage 2 model (observed event rate = 0.38) fit to true exposure, naive exposure, and calibrated exposure with naive (model-based) standard errors,  sandwich standard errors, and  bootstrap standard errors, with bootstrap confidence intervals constructed in 3 ways for the bootstrap samples. We let $N=1000$, $\beta=\log(1.5)$, and vary the correlation between  $X$ and $Z$, the sample size ($N$), and the measurement error variance ($\sigma^2$). Sample size of the calibration subset is $n=450$.}\label{tableBootBCA}
   
        \begin{tabular}{lllcccccccc}

\hline
\multicolumn{2}{l}{} & \multicolumn{4}{c}{Low Correlation}  & \multicolumn{4}{c}{High Correlation}  \\ \cmidrule(r){3-6} \cmidrule{7-10}   $\sigma^2$\tnote{(1)} & Method   & \% Bias & MAD & ASE & CP & \% Bias & MAD & ASE & \multicolumn{1}{c}{CP}   \\ 

\hline
 $0.00$ & Truth & \phantom{0-}0.04 & 0.07 & 0.07 & 0.94 & \phantom{0-}0.16 & 0.09 & 0.10 & 0.94 \\ 
 
\hline
  $0.25$  & Naive &  
  -13.78 & 0.12 & 0.11 & 0.91 & -42.84 & 0.12 & 0.12 & 0.68 \\

    & RC\tnote{(2)} \hspace*{0.2cm} (Naive SE) &     \phantom{0}-3.76 & 0.13 & 0.12 & 0.94 & \phantom{0}-3.28 & 0.21 & 0.20 & 0.94 \\

   & RC\tnote{(2)} \hspace*{0.2cm} (Sandwich) &    \phantom{0}\text{---} & \phantom{0}\text{---}  & 0.12 & 0.94 & \phantom{0}\text{---} & \phantom{0}\text{---} & 0.20 & 0.94 \\ 
 
    & RC\tnote{(2)} \hspace*{0.2cm} (Boot. - Wald)\tnote{(3)} &    \phantom{0}\text{---} &\phantom{0}\text{---}  & 0.12 & 0.94 & \phantom{0}\text{---} & \phantom{0}\text{---} & 0.21 & 0.94 \\ 
  
     & RC\tnote{(2)} \hspace*{0.2cm} (Boot. - Perc)\tnote{(4)} &    \phantom{0}\text{---} &\phantom{0}\text{---}  & 0.12 & 0.93 & \phantom{0}\text{---} & \phantom{0}\text{---}& 0.21 & 0.93 \\

    & RC\tnote{(2)} \hspace*{0.2cm} (Boot. - BCA)\tnote{(5)} &    \phantom{0}\text{---} & \phantom{0}\text{---} & 0.12 & 0.94 & \phantom{0}\text{---} & \phantom{0}\text{---} & 0.21 & 0.94 \\

   \hline
  
 $0.50$  & Naive &  
  -48.54 & 0.08 & 0.08 & 0.36 & -68.36 & 0.09 & 0.09 & 0.13 \\

    & RC\tnote{(2)} \hspace*{0.2cm} (Naive SE) &  \phantom{0}-4.61 & 0.16 & 0.16 & 0.93 &  \phantom{0}-2.90 & 0.28 & 0.27 & 0.94 \\

   & RC\tnote{(2)} \hspace*{0.2cm} (Sandwich) &    \phantom{0}\text{---} &\phantom{0}\text{---}  & 0.16 & 0.93 & \phantom{0}\text{---} & \phantom{0}\text{---} & 0.27 & 0.95 \\ 
 
   & RC\tnote{(2)} \hspace*{0.2cm} (Boot. - Wald)\tnote{(3)} &    \phantom{0}\text{---} &\phantom{0}\text{---}  & 0.16 & 0.94 & \phantom{0}\text{---} & \phantom{0}\text{---} & 0.29 & 0.96 \\

 & RC\tnote{(2)} \hspace*{0.2cm} (Boot. - Perc)\tnote{(4)} &    \phantom{0}\text{---} &\phantom{0}\text{---} &  0.16 & 0.93 & \phantom{0}\text{---} & \phantom{0}\text{---} & 0.29 & 0.94 \\ 

     & RC\tnote{(2)} \hspace*{0.2cm} (Boot - BCA)\tnote{(5)} &    \phantom{0}\text{---} & \phantom{0}\text{---}  & 0.16 & 0.93 & \phantom{0}\text{---} & \phantom{0}\text{---} & 0.29 & 0.94 \\

  \hline
 $1.00$  & Naive &  
 -71.90 & 0.06 & 0.06 & 0.01 & -83.60 & 0.06 & 0.06 & 0.00 \\

    & RC\tnote{(2)} \hspace*{0.2cm} (Naive SE) &    \phantom{-0}-5.31 & 0.22 & 0.21 & 0.94 &  \phantom{0}-4.73 & 0.37 & 0.37 & 0.94 \\ 

    & RC\tnote{(2)} \hspace*{0.2cm} (Sandwich) &    \phantom{0}\text{---} & \phantom{0}\text{---}  & 0.21 & 0.94 & \phantom{0}\text{---} & \phantom{0}\text{---} & 0.37 & 0.95 \\

   & RC\tnote{(2)} \hspace*{0.2cm} (Boot. - Wald)\tnote{(3)}  &    \phantom{0}\text{---} &\phantom{0}\text{---} &   0.22 & 0.96 & \phantom{0}\text{---} & \phantom{0}\text{---} & 0.41 & 0.98 \\

 & RC\tnote{(2)} \hspace*{0.2cm} (Boot. - Perc)\tnote{(4)} &    \phantom{0}\text{---} &\phantom{0}\text{---} &  0.22 & 0.92 & \phantom{0}\text{---} & \phantom{0}\text{---} & 0.41 & 0.94 \\

     & RC\tnote{(2)} \hspace*{0.2cm} (Boot. - BCA)\tnote{(5)} &    \phantom{0}\text{---} &\phantom{0}\text{---} & 0.22 & 0.93 & \phantom{0}\text{---} & \phantom{0}\text{---} & 0.41 & 0.94 \\ 
  \hline

\end{tabular}
 \begin{tablenotes}
     \item[(1)]  $\sigma^2=$ the variance of the random, normally distributed covariate measurement error
      \item[(2)] RC = Regression calibration
    \item[(3)] Bootstrap with standard normal Wald-based confidence interval
      \item[(4)] Percentile bootstrap confidence interval 
      \item[(5)] Bias-corrected and accelerated (BCA) bootstrap interval

   \end{tablenotes}
    \end{threeparttable}
\end{table}

\clearpage

\bibliography{biblio}

\begin{thebibliography}{}

\bibitem[Andersen and Gill, 1982]{andersen1982cox}
Andersen, P.~K. and Gill, R.~D. (1982).
\newblock Cox's regression model for counting processes: a large sample study.
\newblock {\em The annals of statistics}, pages 1100--1120.

\bibitem[Baiocchi et~al., 2014]{baiocchi2014instrumental}
Baiocchi, M., Cheng, J., and Small, D.~S. (2014).
\newblock Instrumental variable methods for causal inference.
\newblock {\em Statistics in Medicine}, 33(13):2297--2340.

\bibitem[Baldoni et~al., 2021]{baldoni2021use}
Baldoni, P.~L., Sotres-Alvarez, D., Lumley, T., and Shaw, P.~A. (2021).
\newblock On the use of regression calibration in a complex sampling design
  with application to the hispanic community health study/study of latinos.
\newblock {\em American Journal of Epidemiology}.

\bibitem[Baron and Kenny, 1986]{baron1986moderator}
Baron, R.~M. and Kenny, D.~A. (1986).
\newblock The moderator--mediator variable distinction in social psychological
  research: Conceptual, strategic, and statistical considerations.
\newblock {\em Journal of Personality and Social Psychology}, 51(6):1173.

\bibitem[Binder, 1983]{binder1983variances}
Binder, D.~A. (1983).
\newblock On the variances of asymptotically normal estimators from complex
  surveys.
\newblock {\em International Statistical Review/Revue Internationale de
  Statistique}, pages 279--292.

\bibitem[Boos and Stefanski, 2013]{boos2013essential}
Boos, D. and Stefanski, L. (2013).
\newblock {\em Essential Statistical Inference: Theory and Methods}.
\newblock Springer, New York, NY.

\bibitem[Davison and Hinkley, 1997]{davison1997bootstrap}
Davison, A.~C. and Hinkley, D.~V. (1997).
\newblock {\em Bootstrap methods and their application}.
\newblock Number~1. Cambridge university press.

\bibitem[Efron, 1979]{efron1979bootstrap}
Efron, B. (1979).
\newblock Bootstrap methods: Another look at the jackknife.
\newblock {\em The Annals of Statistics}, 7(1):1--26.

\bibitem[Efron, 1987]{efron1987better}
Efron, B. (1987).
\newblock Better bootstrap confidence intervals.
\newblock {\em Journal of the American Statistical Association},
  82(397):171--185.

\bibitem[Foutz, 1977]{foutz1977unique}
Foutz, R.~V. (1977).
\newblock On the unique consistent solution to the likelihood equations.
\newblock {\em Journal of the American Statistical Association},
  72(357):147--148.

\bibitem[{Hispanic Community Health Study/Study of Latinos}, 2020]{SOLdatacite}
{Hispanic Community Health Study/Study of Latinos} (2020).
\newblock {HCHS/SOL Investigator Datasets}.
\newblock \url{https://sites.cscc.unc.edu/hchs/}.

\bibitem[Keogh et~al., 2020]{keogh20}
Keogh, R.~H., Shaw, P.~A., Gustafson, P., Carroll, R.~J., Deffner, V., Dodd,
  K.~W., K{\"u}chenhoff, H., Tooze, J.~A., Wallace, M.~P., Kipnis, V., and
  Freedman, L. (2020).
\newblock Stratos guidance document on measurement error and misclassification
  of variables in observational epidemiology: Part 1—basic theory and simple
  methods of adjustment.
\newblock {\em Statistics in Medicine}, 39(16):2197--2231.

\bibitem[Kipnis et~al., 2009]{kipnis2009modeling}
Kipnis, V., Midthune, D., Buckman, D.~W., Dodd, K.~W., Guenther, P.~M.,
  Krebs-Smith, S.~M., Subar, A.~F., Tooze, J.~A., Carroll, R.~J., and Freedman,
  L.~S. (2009).
\newblock Modeling data with excess zeros and measurement error: application to
  evaluating relationships between episodically consumed foods and health
  outcomes.
\newblock {\em Biometrics}, 65(4):1003--1010.

\bibitem[Langer et~al., 2003]{langer2003women}
Langer, R.~D., White, E., Lewis, C.~E., Kotchen, J.~M., Hendrix, S.~L., and
  Trevisan, M. (2003).
\newblock {The Women's Health Initiative Observational Study: baseline
  characteristics of participants and reliability of baseline measures}.
\newblock {\em Annals of Epidemiology}, 13(9):S107--S121.

\bibitem[Lumley, 2011]{lumley2011complex}
Lumley, T. (2011).
\newblock {\em Complex surveys: a guide to analysis using R}, volume 565.
\newblock John Wiley \& Sons.

\bibitem[Lumley and Scott, 2017]{lumley2017fitting}
Lumley, T. and Scott, A. (2017).
\newblock Fitting regression models to survey data.
\newblock {\em Statistical Science}, pages 265--278.

\bibitem[Mossavar-Rahmani et~al., 2015]{mossavar2015applying}
Mossavar-Rahmani, Y., Shaw, P.~A., Wong, W.~W., Sotres-Alvarez, D., Gellman,
  M.~D., Van~Horn, L., Stoutenberg, M., Daviglus, M.~L., Wylie-Rosett, J.,
  Siega-Riz, A.~M., et~al. (2015).
\newblock Applying recovery biomarkers to calibrate self-report measures of
  energy and protein in the hispanic community health study/study of latinos.
\newblock {\em American Journal of Epidemiology}, 181(12):996--1007.

\bibitem[Neuhouser et~al., 2008]{neuhouser2008use}
Neuhouser, M.~L., Tinker, L., Shaw, P.~A., Schoeller, D., Bingham, S.~A., Horn,
  L.~V., Beresford, S.~A., Caan, B., Thomson, C., Satterfield, S., et~al.
  (2008).
\newblock {Use of recovery biomarkers to calibrate nutrient consumption
  self-reports in the Women's Health Initiative}.
\newblock {\em American Journal of Epidemiology}, 167(10):1247--1259.

\bibitem[Prentice, 1982]{prentice1982covariate}
Prentice, R.~L. (1982).
\newblock Covariate measurement errors and parameter estimation in a failure
  time regression model.
\newblock {\em Biometrika}, 69(2):331--342.

\bibitem[Prentice et~al., 2021]{prentice2021biomarker}
Prentice, R.~L., Pettinger, M., Neuhouser, M.~L., Raftery, D., Zheng, C.,
  Gowda, G.~N., Huang, Y., Tinker, L.~F., Howard, B.~V., Manson, J.~E., et~al.
  (2021).
\newblock Biomarker-calibrated macronutrient intake and chronic disease risk
  among postmenopausal women.
\newblock {\em The Journal of Nutrition}, 151(8):2330--2341.

\bibitem[Prentice et~al., 2022]{prentice2022biomarkers}
Prentice, R.~L., Pettinger, M., Zheng, C., Neuhouser, M.~L., Raftery, D.,
  Gowda, G.~N., Huang, Y., Tinker, L.~F., Howard, B.~V., Manson, J.~E., et~al.
  (2022).
\newblock Biomarkers for components of dietary protein and carbohydrate with
  application to chronic disease risk in postmenopausal women.
\newblock {\em The Journal of Nutrition}, 152(4):1107--1117.

\bibitem[Ritenbaugh et~al., 2003]{ritenbaugh2003women}
Ritenbaugh, C., Patterson, R.~E., Chlebowski, R.~T., Caan, B., Fels-Tinker, L.,
  Howard, B., and Ockene, J. (2003).
\newblock {The Women's Health Initiative Dietary Modification trial: overview
  and baseline characteristics of participants}.
\newblock {\em Annals of Epidemiology}, 13(9):S87--S97.

\bibitem[S{\"a}rndal et~al., 2003]{sarndal2003model}
S{\"a}rndal, C.-E., Swensson, B., and Wretman, J. (2003).
\newblock {\em Model assisted survey sampling}.
\newblock Springer Science \& Business Media.

\bibitem[Shaw et~al., 2018]{shaw2018citation}
Shaw, P.~A., Deffner, V., Keogh, R.~H., Tooze, J.~A., Dodd, K.~W.,
  K{\"u}chenhoff, H., Kipnis, V., and Freedman, L.~S. (2018).
\newblock Epidemiologic analyses with error-prone exposures: Review of current
  practice and recommendations.
\newblock {\em Annals of Epidemiology}, 27(11):821--828.

\bibitem[{The Women's Health Initiative Study Group}, 1998]{study1998design}
{The Women's Health Initiative Study Group} (1998).
\newblock {Design of the Women’s Health Initiative clinical trial and
  observational study}.
\newblock {\em Controlled Clinical Trials}, 19(1):61--109.

\bibitem[Tinker et~al., 2011]{tinker2011biomarker}
Tinker, L.~F., Sarto, G.~E., Howard, B.~V., Huang, Y., Neuhouser, M.~L., et~al.
  (2011).
\newblock {Biomarker-calibrated dietary energy and protein intake associations
  with diabetes risk among postmenopausal women from the Women's Health
  Initiative}.
\newblock {\em The American Journal of Clinical Nutrition}, 94(6):1600--1606.

\bibitem[Van~der Vaart, 2000]{van2000asymptotic}
Van~der Vaart, A.~W. (2000).
\newblock {\em Asymptotic statistics}, volume~3.
\newblock Cambridge University Press.

\bibitem[{Women's Health Initiative}, 2019]{WHIdatacite}
{Women's Health Initiative} (2019).
\newblock Women's {H}ealth {I}nitiative {I}nvestigator {D}atasets.
\newblock \url{https://www.whi.org}.

\end{thebibliography}

\clearpage

\end{document}